\documentclass[10pt]{article}
\usepackage{geometry}
\usepackage{xcolor}
\definecolor{graycolor}{gray}{0.9} 
\usepackage{microtype}
\usepackage{setspace} 
\usepackage[utf8]{inputenc}
\usepackage[english]{babel}
\usepackage{times}
\usepackage{array}
\usepackage{soul}
\usepackage{cite}
\usepackage[numbers,sort&compress]{natbib}
\usepackage{natbib}

\usepackage{titlesec} 
\titleformat {\section} [block] {\raggedright \fontsize{10}{10}\selectfont\bfseries} {\thesection. \space} {0pt} {}
\titlespacing {\section} {0pt} {12pt} {6pt}
\titleformat {\subsection} [block] {\raggedright \fontsize{10}{10}\selectfont\itshape} {\thesubsection .\space} {0pt} {}
\titlespacing {\subsection} {0pt} {12pt} {6pt}
\titleformat {\subsubsection} [block] {\raggedright \fontsize{10}{10}\selectfont} {\thesubsubsection .\space} {0pt} {}
\titlespacing {\subsubsection} {0pt} {12pt} {6pt}
\titleformat {\paragraph} [block] {\raggedright \fontsize{10}{10}\selectfont} {} {0pt} {}
\titlespacing {\paragraph} {0pt} {12pt} {6pt}

\usepackage{array} \newcommand{\PreserveBackslash}[1]{\let\temp=\\#1\let\\=\temp}
\newcolumntype{C}[1]{>{\PreserveBackslash\centering}m{#1}}
\newcolumntype{R}[1]{>{\PreserveBackslash\raggedleft}m{#1}}
\newcolumntype{L}[1]{>{\PreserveBackslash\raggedright}m{#1}}
\usepackage{lineno}
\usepackage{tabularx}
\usepackage{colortbl}
\usepackage{graphicx}
\usepackage{float}
\usepackage[export]{adjustbox}
\usepackage{caption}
\usepackage{fancyhdr} 
\usepackage{lastpage}
\usepackage{layout}
\usepackage{setspace} 
\usepackage{enumitem}
\usepackage{booktabs}
\usepackage{arydshln}
\usepackage{multirow}
\usepackage{color}
\usepackage{hyperref} 
\hypersetup{
	colorlinks=true,
	linkcolor=blue,
	filecolor=blue,
	urlcolor=black,
	citecolor=cyan,
}

\setcitestyle{open={[},close={]},citesep={,\!},numbers}
\soulregister\cite7 
\soulregister\ref7

\fancypagestyle{firstpage}{
    \setlength{\headsep}{2.2cm}
    
    \setlength{\footskip}{1.5cm}
    \fancyhf{}
    \lhead{\begin{table}[H]
        \centering
        \begin{tabular}{L{2.5cm}C{10cm}C{3.1cm}R{2cm}}
            \includegraphics[scale=0.035]{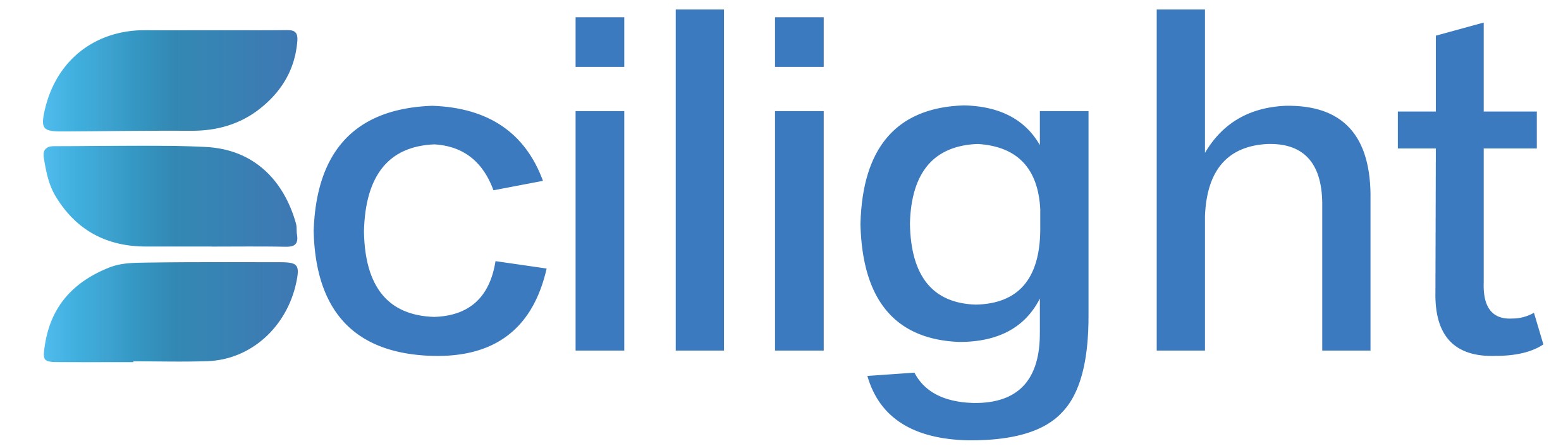} \vspace{-6pt}& \cellcolor{graycolor}\begin{tabular}[c]{@{}c@{}}\textit{Pragmatic Cybersecurity}\\ \href{https://www.sciltp.com/journals/pc}{https://www.sciltp.com/journals/pc}\end{tabular} & \includegraphics[scale=0.022]{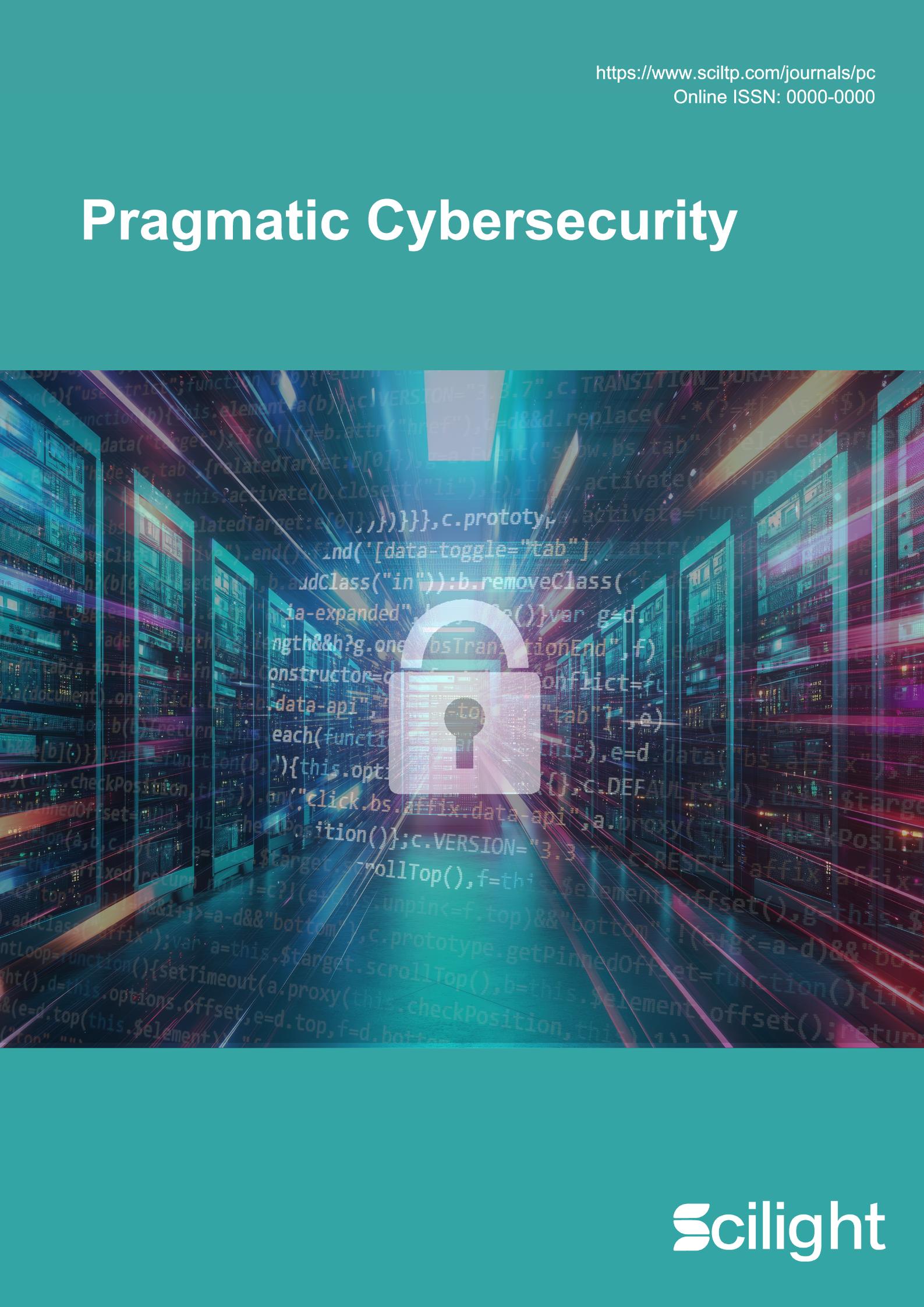} \vspace{-3pt}\\
        \end{tabular}
        \vspace{-22pt}
    \end{table}}
   
    \fancyfoot[C]{
        \vspace{-1.55cm}
        \begin{table}[H]
            \begin{minipage}[c]{0.15\columnwidth}
                \includegraphics[scale=0.5]{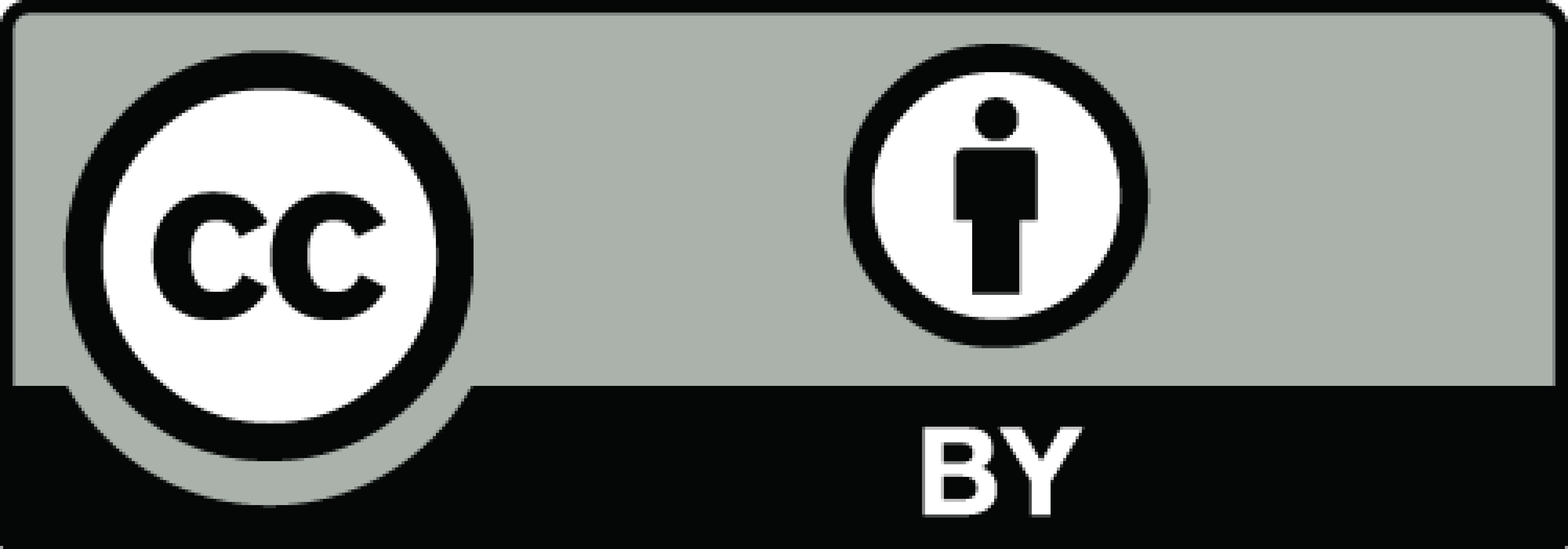} \vspace{1.1pt}
            \end{minipage}
            \hfill
            \begin{minipage}[c]{0.85\columnwidth}
                \scriptsize \textbf{Copyright:} © 2026 by the authors. This is an open access article under the terms and conditions of the Creative Commons Attribution (\mbox{CC BY}) license (\href{https://creativecommons.org/licenses/by/4.0/}{https://creativecommons.org/licenses/by/4.0/}). \\ \textbf{Publisher’s Note:} Scilight stays neutral with regard to jurisdictional claims in published maps and institutional affiliations.
            \end{minipage}
    \end{table}}
    \vspace{-0.55cm}
}

\begin{document}
\newgeometry{left=2.5cm, right=2.5cm, top=1.8cm, bottom=4cm}
	\thispagestyle{firstpage}
	{\noindent \textit{Article}}
	\vspace{4pt} \\
	{\raggedright\fontsize{18pt}{10pt}\textbf{Global Web, Local Privacy? An International Review of Web Tracking}\par }\vspace{16pt}
	{\raggedright\large Harry Yu \textsuperscript{1,\textdagger,\textdaggerdbl}, Patton Yin \textsuperscript{2,\textdagger,\textdaggerdbl} and Sebastian Zimmeck \textsuperscript{3,*} }
	\vspace{6pt}
	 \begin{spacing}{0.9}
		{\noindent \small
			\textsuperscript{1}	\parbox[t]{0.98\linewidth}{Department of Computer Science, Carnegie Mellon University, Pittsburgh, PA 15213, United States}\\ 
			\textsuperscript{2}	Department of Computer Science, Brown University, Providence, RI 02912, United States \\
			\textsuperscript{3}	Department of Mathematics and Computer Science, Wesleyan University, Middletown, CT 06459, United States \\
		    {*}  \parbox[t]{0.98\linewidth}{Correspondence: szimmeck@wesleyan.edu} \\
			{\textdagger}  These authors contributed equally to this work. \\
			{\textdaggerdbl}  They performed their work during their studies at Wesleyan University. 
 			\vspace{6pt}\\
		\footnotesize	\textbf{How To Cite}: Yu, H.; Yin, P.; Zimmeck, S. Global Web, Local Privacy? An International Review of Web Tracking. \emph{Pragmatic Cybersecurity} \textbf{2026}, \emph{1}(1), 5. }\\
	\end{spacing}

\hl{Preliminary Draft. See https://www.sciltp.com/journals/pc/articles/2603003347 for the final version.}

\begin{table}[H]
\noindent\rule[0.15\baselineskip]{\textwidth}{0.5pt} 
\begin{tabular}{lp{12cm}}  
 \small 
  \begin{tabular}[t]{@{}l@{}} 
  \footnotesize  Received: 31 December 2025 \\
  \footnotesize  Revised: 11 March 2026 \\
   \footnotesize Accepted: 17 March 2026 \\
  \footnotesize  Published: day month year
  \end{tabular} &
  \textbf{Abstract:} Web tracking by ad networks, social networks, and other third parties is privacy-invasive. To protect users' privacy an increasing number of countries are adopting new privacy laws.
  However, a major reason why their application on the web is so challenging is that privacy laws are local while the web is global.
  To that end, we evaluate websites' tracker connections for ten countries for two sets of sites---the global Common Top 525 and the Country-specific Top 525 sites. 
  We find that Australia and the US (California)---two of the three opt-out jurisdictions in our study---have the highest level of web tracking while opt-in jurisdictions generally have lower levels.
  We also find that the Common Top 525 sites have 50.5\% fewer average tracker connections when accessed from EU countries compared to non-EU countries. 
  Further, simply not interacting with cookie banners decreases trackers by 48.5\% for Germany, as measured for a sample of 36 Common Top 525 sites.
  These results suggest that the General Data Protection Regulation and the ePrivacy Directive have a tangible effect in reducing tracking.
  As 28\% of Common Top 525 sites show cookie banners in all ten countries, our results suggest a moderate Brussels effect.
  However, against the backdrop of global US ad tech practices, EU law primarily acts as a Brussels shield.
  Generally, we think that strong enforcement of privacy laws is key to increase user privacy on the web. \\
  & 
\end{tabular}
\noindent\rule[0.15\baselineskip]{\textwidth}{0.5pt} 
\end{table}

	\section{Introduction}
\label{Introduction}


The most common business model for the web is ``content for data
.'' 
In exchange for access to news, videos, games, and other online content, users pay with their data.
Ad networks and other third parties integrated on websites track users' online activities---for example, what they like, what they purchase, or where they are located---to create user profiles, deliver targeted ads, and personalize their web experience~\cite{Weinshel2019CCS}. 
HTTP cookies, for example, used in combination with invisible tracking pixels, are some of the most common tracking mechanisms. 
Their use for tracking has long been recognized as privacy-invasive, and increasingly lawmakers and regulators aim to curb the harm of web tracking with privacy laws~\cite{7872467} (For simplicity we refer to ``privacy laws'' in this study to cover both ``privacy laws,'' such as the California Consumer Privacy Act (CCPA), and ``data protection laws,'' such as the General Data Protection Regulation (GDPR). Similarly, we use other broad terms to cover technical terms under specific laws, e.g., ``user'' covers `consumer'' under the CCPA and ``data subject'' under the GDPR). 
Over the past decade lawmaking activity has picked up substantially.
An increasing number of countries enacted comprehensive privacy laws that establish which personal data website operators and third parties can collect and under which conditions and with whom they can share it. 
Once these new laws are in place, regulators are being tasked with enforcing them.
Despite this lawmaking activity, questions remain how effective privacy laws are in practice.

\restoregeometry 

In order to protect users from privacy-invasive tracking on websites, two common approaches are: (1)~prohibiting tracking by default, while allowing users to opt in and (2) allowing tracking by default and giving user a right to opt out.
The EU's GDPR~\cite{GDPR2016} and ePrivacy Directive~\cite{directive2009_136} are based on the former and so is Brazil's Lei Geral de Proteção de Dados (LGPD)~\cite{capes_lgpd_overview} as well as South Korea's Personal Information Protection Act (PIPA)~\cite{pipa_korea_2023}, among others. 
In contrast, opt-out laws---most notably the CCPA and other state privacy laws in the US~\cite{CCPA_Overview}---allow tracking by default, thus, requiring users to take action if they wish to restrict tracking. 
As long as users in opt-in jurisdictions do not opt in, they should be tracked less than users in opt-out jurisdictions.
However, they may be subjected to a deluge of cookie banners and experience consent fatigue~\cite{CHOI201842}.

A major reason why the application of privacy laws on the web is so challenging is that privacy laws are local while the web is global.
Different countries and regions have different privacy laws, and many have none.
Different countries are also at different stages in their evolution of privacy lawmaking and enforcement. 
Some laws may also create spillover effects impacting regions beyond their territorial scope, for example, as website operators may decide to simplify their compliance by adhering to the strictest law that applies to their site.
One notable example is the Brussels effect~\cite{BrusselsEffect}.
Overall, web users' levels of privacy can differ depending on geographic locations.
Thus, we are evaluating trackers across regions to understand how different privacy laws shape tracking and consent practices locally and globally.

\subsection{Research Questions}
\label{Research-Questions}

With this study we are addressing the following research questions:

\begin{enumerate}[label=\textbf{RQ\arabic*.},topsep=3pt,parsep=0pt,itemsep=0pt,leftmargin=*,labelsep=0.5mm,align=parleft] 
    \item \textls[-15]{Sites Across Countries 
: How do users' geographic locations in different countries impact their tracker exposure?}
    \item Global vs. Local Sites: Do globally popular sites have different tracking practices than locally popular sites?
    \item Potential Privacy Law Impact: How are sites' tracking and consent practices shaped by privacy laws across countries?
\end{enumerate}

\subsection{General Approach}
\label{General Approach}

We address our research questions by measuring and evaluating the tracker connections of websites in ten  countries: Australia, Brazil, Canada, Germany, India, Singapore, South Africa, South Korea, Spain, and the US.
The ten countries cover all continents except Antarctica.
For each country, we evaluate two sets of sites: the Common Top 525 sites---a set of globally popular sites---and the Country-specific Top 525 sites---a set of popular sites in each of the evaluated countries.
Website operators can ensure compliance with three primary methods: (1) not issuing tracking requests, (2) blocking data collection until obtaining user consent, or (3) allowing data collection accompanied by an indicator of the user's consent status.
We focus here on the first method because it is directly observable and comparable across all site visits without requiring site-specific interactions or assumptions about third party behavior.
While sites might connect to third parties and rely on them to discard data server-side, the absence of a connection guarantees that no data was shared.

\subsection{Contributions}
\label{Contributions}

In the following, we will begin our inquiry with a discussion of related work (Section~\ref{Related-Work}).
Based on the synthesis of applicable privacy laws (Section~\ref{Privacy-Laws}), we then perform a privacy evaluation of websites' tracker connections in ten different countries (Sections~\ref{Methodology} and~\ref{Results}) that we contextualize for the broader web ecosystem to improve its overall privacy (Section~\ref{Discussion}) (The code of our web crawler and data related to this study is publicly available under the MIT License~\cite{privacyPioneerWebCrawler}).

\section{Related Work}
\label{Related-Work}

We see our study as a contribution to the field of web privacy, especially to measurement studies with multi-country scope that evaluate the influence of the GDPR and other privacy laws on cookie consent, compliance, and how users exercise their privacy rights.

\subsection{Web Privacy Measurement}

Various studies have shown how third-party trackers use cookies, browser fingerprinting, and other privacy-invasive technologies to follow users in their browsing sessions across sites~\cite{10.1145/2508859.2516674, 10.1145/2976749.2978313, 10.1145/2736277.2741679}. 
An early study in 2012 showed that most commercial sites implemented trackers by multiple parties, with some capturing over 20\% of a user's browsing behavior~\cite{10.5555/2228298.2228315}. 
A longitudinal study showed that from 1996 to 2016, web tracking grew in prevalence and complexity~\cite{197133}. 
This research has been enabled by measurement platforms like FourthParty~\cite{6234427} and OpenWPM~\cite{10.1145/2976749.2978313} as well as browser extensions like Disconnect~\cite{disconnect_tracker_protection} and Privacy Pioneer~\cite{zimmeckEtAlPrivacyPioneer2024}, the latter of which we use here as well.
As privacy law-making activity picked up substantially in the late 2010s various studies aimed to quantify the impact of the GDPR and other privacy laws on web tracking.
Longitudinal studies comparing sites' privacy practices before and after the time the GDPR went into effect showed an increase in cookie consent notices~\cite{Degeling_2019} as well as a decline in the number of third parties belonging to certain categories, though, it remained unclear whether the latter was attributable to the GDPR~\cite{10.1145/3308558.3313524}.

\subsection{Impact of the GDPR and ePrivacy Directive}

The GDPR aims to improve data protection online, which leads to a direct conflict with the online ad industry's real-time bidding infrastructure for ads personalization~\cite{Veale_Zuiderveen_Borgesius_2022}.
In some areas the law's impact has improved data protection on the web while in others its full impact has yet to be realized. 
Studies have shown that after the GDPR's enforcement, sites embedded fewer trackers~\cite{solomos2020clashtrackersmeasuringevolution} and reduced connections to web technology providers, which was also the case for sites intended for non-EU audiences~\cite{EuropeanPrivacyLawandGlobalMarketsforData}. 
The number of ID syncing connections also decreased, reducing information sharing between third parties~\cite{GDPRImpact}. 
However, the overall structure of the tracking ecosystem remains largely unaffected, and the GDPR may have inadvertently increased market concentration in the online ad industry, with fewer, larger companies dominating web tracking~\cite{EuropeanPrivacyLawandGlobalMarketsforData,GDPRImpact}. 
An early study covering the Netherlands' implementation of the ePrivacy Directive into national law concluded that it did not result in meaningful choice for users but instead caused widespread deployment of annoying banners, pop-up screens, and cookie walls~\cite{LEENES2015317}.

Beyond the EU, the introduction of the GDPR has increased friction between jurisdictions, which is most evident in transatlantic data flows. 
The 2020 Schrems II ruling by the Court of Justice of the European Union~\cite{SchremsII}, which invalidated the EU-US Privacy Shield, concluded that US law does not provide ``essentially equivalent'' protection for EU residents' data, thereby complicating third-party tracking that relies on US-based servers~\cite{Rubinstein}. 
Despite these hurdles, the GDPR often functions as a global de facto standard as multinational companies often apply EU-level protections across their global operations to minimize technical and legal fragmentation~\cite{10.1093/idpl/ipaf005}. 
However, work comparing EU and US cookie behavior suggests that while persistent tracking has decreased in the EU, US-based sites continue to maintain significantly higher tracking densities due to the permissive opt-out nature of US privacy laws~\cite{10.1007/978-3-030-15986-3_17}.

\subsection{Cookie Consent, Compliance, and Opting Out}

A large body of research demonstrates widespread non-compliance with the consent requirements of the ePrivacy Directive and GDPR. 
Studies have consistently found that sites install tracking cookies before receiving user consent~\cite{4YearsofEUCookieLawResultsandLessonsLearned}.
Even when consent is sought, the mechanisms are often flawed. 
Sites deploy consent banners with deceptive designs that nudge users towards acceptance or assume consent from inaction~\cite{CookieBannersIAB}. Many sites continue to collect data even when users explicitly reject consent~\cite{bollinger2022automating, BouhoulaAutomatedLargeScaleCookieNoticeCompliance}. 
The rise of the ``consent ecosystem,'' dominated by a few Consent Management Platforms (CMPs), has led to standardized but often confusing and unusable consent dialogs that encourage users to ``Accept All''~\cite{HilsMeasuringEmergenceOfConsentManagementOnTheWeb, DBLP:journals/corr/abs-1908-10048}, have no reject option~\cite{10.1145/3706598.3713648}, or include dark patterns and implied consent features~\cite{10.1145/3313831.3376321, 10.1145/3706598.3713648}. 
Users also often experience consent fatigue~\cite{CHOI201842}.
The automation of rejecting consent can help to tip the scale towards user privacy~\cite{263860}.
Privacy preference signals, such as Global Privacy Control (GPC)~\cite{hausladenEtAlGPCWeb2025,zimmeckEtAlGPC2023}, can mitigate this impact~\cite{rasaii2025intractablecookiecrumbsunveiling}, however, have yet to be adopted broadly. 
Furthermore, it is a fundamental design flaw of the online ad ecosystem that it is built on the notion that sites make normal ad calls to third parties, append opt-out flags, for example of the Transparency Consent Framework~\cite{IABTransparencyConsentFramework}, and leave it to the downstream third parties to respect the opt-outs by discarding the received user data~\cite{zimmeck2021connections}.

\subsection{Multi-country Privacy Law Comparisons}

Previous studies highlight significant differences in privacy practices across jurisdictions. 
A key finding is that cookie behavior, cookie consent violation rates, and cookie banner implementations are highly dependent on region~\cite{tang2025navigatingcookieconsentviolations}.
Sites employ fewer cookies for EU visitors compared to US visitors~\cite{10.1007/978-3-030-15986-3_17}.
While the Brussels effect has been observed, i.e., EU standards being applied to sites' US visitors~\cite{10.1007/978-3-030-15986-3_17}, the impact of EU law on the operations of US online services is limited~\cite{Frankenreiter}.
Cookie banner existence even decreased when
accessing European websites from the US~\cite{294572}. 
Even within the EU, the transition to the GDPR saw inconsistent adoption of functional and usable consent mechanisms across member states~\cite{Degeling_2019}.
Sites appear to follow notice requirements based
on their expected audience, as identified by their country code top-level domain, and not individual user location, except for \texttt{.com} sites~\cite{Eijk2019TheIO}.
These implementation differences are mirrored by cultural variations that influence privacy concerns and the application of legal principles like transparency across borders~\cite{InternationalDifferences,TheRoleOf,CulturalDifferences,10.1093/idpl/ipae011}. 
To overcome the challenges of privacy self-management researchers have called for the study of alternative approaches~\cite{10646758}.
In particular, laws can introduce structural measures that would relieve individuals from unrealistic privacy self-management expectations~\cite{SolovePrivacyRights}.

\section{Privacy Laws and their Enforcement}
\label{Privacy-Laws}

We study websites in ten countries: Australia, Brazil, Canada, Germany, India, Singapore, South Africa, South Korea, Spain, and the US (California).
The territorial scope of the privacy laws in these countries, whether they require opt-in or opt-out consent, and their enforcement determine their applicability and impact.

\subsection{Territorial Scope}

Typically, website operators and third parties are subject to the privacy laws of their jurisdiction.
However, many laws also have extraterritorial application.
For example, the GDPR not only applies to controllers and processors established in the EU but also to those outside it if they offer goods or services to, or monitor the behavior of, data subjects within the EU (GDPR, Article 3(2)).
Similarly, the ePrivacy Directive applies to the processing of personal data in public communications networks in the European Community, i.e., EU, regardless of who does the processing (ePrivacy Directive, Article 3(1)).
In the US, the CCPA applies to any for-profit entity that does business in the State of California and that has annual gross revenues in excess of twenty-five million dollars, annually buys, sells, or shares the personal information of 100,000 or more consumers or households, or derives 50 percent or more of its annual revenues from selling or sharing consumers' personal information (CCPA, Section 1798.140(d)(1)).
Generally, all privacy laws in our study protect users within their jurisdiction regardless of where the data processing takes place (See the Australian Privacy Act 1988, Article 5B (1A) and (3), Brazil's LGPD, Article 3, India's Digital Personal Data Protection (DPDP) Act, 2023, Article 3, and 
South Africa's Protection of Personal Information Act (POPIA), Section 3. While Canada's Personal Information Protection and Electronic Documents Act (PIPEDA) is silent on the matter, courts interpret the law to apply where there is a ``real and substantial link'' to Canada~\cite{PIPEDATerritorialScope}. Similar guidance is provided by governmental regulators in Singapore~\cite{SingaporeTerritorialScope} and South Korea~\cite{SouthKoreaTerritorialScope}).
We ensure applicability of a jurisdiction's privacy law by accessing sites from a Virtual Machine (VM) server located inside it.

\subsection{Consent Type}

Most privacy laws require user consent for web tracking; many in form of opt-in consent and some in form of opt-out consent.
Opt-in laws require website operators to obtain consent from users before collecting any personal data or sharing it with third parties, such as in the case of non-essential cookies for advertising or analytics.
Users in opt-out jurisdictions, on the other hand, can be tracked until they exercise their right to opt out.
Thus, without any user interaction we expect lower levels of tracking in opt-in jurisdictions (By ``opt-in jurisdiction'' and ``opt-out jurisdiction'' we refer to the type of \textit{consent} a user is required to give. We do not mean to refer to other legal bases for personal data processing, for example, legitimate interest per GDPR, Article 6(1)(f). Under EU law we focus here on consent because the Court of Justice of the European Union held that a user ``cannot reasonably expect that the operator [\ldots] will process that user's personal data, without his or her \textit{consent}, for the purposes of personalised advertising'' (emphasis added)~\cite{MetaVsBundeskartellamt}. Furthermore, the ePrivacy Directive's requirement for prior consent per Article 5(3) applies to the access of information on a user-s device---a technical prerequisite for web tracking---thereby requiring an opt-in consent before any data collection can occur~\cite{Planet49, 10.1093/idpl/ipv011}).

\subsubsection{Opt-In Jurisdictions}

Germany, Spain, Brazil, India, South Korea, Singapore, and South Africa 
 are opt-in jurisdictions.
For Germany and Spain, the ePrivacy Directive requires consent as the legal basis for all cookies that are not strictly necessary for the operation of a website (ePrivacy Directive, Article 5(3)).
Access to user information is only allowed on the condition that the user is provided with clear and comprehensive information about the purposes of the processing and is offered the right to refuse such processing.
The Court of Justice of the European Union further specified the law in its Planet49 decision reasoning that pre-checked boxes are insufficient and that users must be notified how long cookies will be stored on their devices and whether or not third parties may have access to those~\cite{Planet49}. 

For Brazil the LGPD enumerates the legal bases for processing personal data in Article 7 and requires user consent as the default basis for processing personal data via tracking cookies.
Brazil's National Data Protection Authority published guidance against using cookie banners with pre-selected authorization options or the adoption of tacit consent mechanisms, such that by continuing to browse a site a user is assumed to give consent~\cite{BrazilCookieGuidance}.
This is also true for India, whose DPDP Act, 2023 requires consent to be ``unambiguous with a clear affirmative action'' (DPDP Act, 2023, Article 6(1)).
Similarly, opt-in consent is required per Article 22(1)(No.7) of South Korea's PIPA as well as per regulatory guidance in Singapore~\cite{SingaporeOptInGuidance} and South Africa~\cite{SouthAfricaOptInGuidance}.

\subsubsection{Opt-Out Jurisdictions}

Australia, Canada, and the US (California) are opt-out jurisdictions.
The Office of the Australian Information Commissioner issued guidance on the use of tracking pixels according to which organizations should enable users to opt-out of receiving targeted online ads using tracking pixels, for example, by deploying a banner or pop-up when a user first visits a site~\cite{AustraliaEnforcement}.
Similarly, the Privacy Commissioner of Canada issued guidance on the use of opt-out consent for online behavioral advertising requiring that users are notified and no sensitive information is involved~\cite{PIPEDAGuidance}.
In the US the CCPA operates on an opt-out basis for the selling and sharing of personal information, the latter of which includes communicating a consumer's personal information to a third party for cross-context behavioral advertising (CCPA, Section 1798.140(ah)(1)).
Websites must provide a ``Do Not Sell or Share My Personal Information'' link and honor GPC opt-out preference signals~\cite{CCPAEnforcementCaseExamples}. 

\subsection{Privacy Law Enforcement}
\label{Privacy-Law-Enforcement}

For enforcing users' privacy rights, especially, consent choices, we distinguish jurisdictions with high, medium, and low enforcement activity (Table~\ref{tab:privacy-country-law-overview}).

\vspace{-16pt}
\begin{table}[H]
\caption{Germany and Spain are the only countries in our study with high enforcement activity.}
\resizebox{\columnwidth}{!}{
\centering
\small
\begin{tabular}{lllll}
\toprule
\textbf{Country} & \textbf{Consent Type} & \textbf{Enforcement} & \textbf{Major Enforcement Actions} & \textbf{Primary Privacy Law} \\
\midrule
Germany \& Spain      & Opt-in  & High   & EU-wide 833 fines for insufficient legal basis & GDPR, ePrivacy Directive \\
      &   &    & for data processing (\texteuro 3,011,611,435) & \\
Australia     & Opt-out & Medium & Meta Cambridge Analytica settlement (\$50,000,000) & Privacy Act 1988 \\
Canada        & Opt-out & Medium & Meta Cambridge Analytica investigation & PIPEDA \\
South Korea   & Opt-in  & Medium & Meta investigation (approximately \$15,000,000)   & PIPA \\
US (California) & Opt-out & Medium & Individual enforcement actions by the California & CCPA \\
 & & & Attorney General and the CPPA &  \\
Brazil        & Opt-in  & Low    & Sanctions mostly for data breaches, not tracking & LGPD \\
India         & Opt-in  & Low    & No major enforcement actions & DPDP Act, 2023 \\
Singapore     & Opt-in  & Low    & No major enforcement actions & PDPA \\
South Africa  & Opt-in  & Low    & No major enforcement actions & POPIA \\
\bottomrule
\end{tabular}}
\label{tab:privacy-country-law-overview}
\end{table}

\subsubsection{High Enforcement Activity}

Germany and Spain, which represent the EU jurisdictions in our study, are high enforcement jurisdictions.
They have a strong and active regulatory environment related to tracking and user consent.
Especially, Germany has a long privacy law and enforcement tradition dating back to the 1970s. 
The German Data Protection Authorities are among the most active in the EU and enforce privacy laws broadly.
For example, in a recent case brought by the Data Protection Authority of Lower Saxony against a web publisher the Administrative Court Hannover decided that the use of the options ``Accept all,'' ``Accept \& close x,'' and ``Settings'' is insufficient for valid consent and that the use of Google Tag Manager requires consent~\cite{VGHannover}.
The Spanish Data Protection Authority is also very active and recently fined the automotive company SEAT for placing non-technical cookies without user consent and failing to stop such placement after consent withdrawal~\cite{AEPD}.
As of March 2026, the EU has issued a total of 833 fines against companies on the ground of insufficient legal basis for data processing for a total of \texteuro 3,011,611,435~\cite{GDPREnforcementTracker}.

\subsubsection{Medium Enforcement Activity}

Australia, Canada, South Korea, and the US (California), also have engaged in enforcing applicable privacy laws.
However, their enforcement activities are generally not broad-based but instead focused on individual high-profile cases.
We categorize them as medium enforcement jurisdictions.
The Office of the Australian Information Commissioner issued guidance on the use of tracking pixels~\cite{AustraliaEnforcement} and settled a civil case with Meta on the sharing of user data with Cambridge Analytica for \$50,000,000~\cite{AustraliaEnforcementMeta}.
The Privacy Commissioner of Canada and the Information and Privacy Commissioner for British Columbia also investigated Meta and found various consent shortcomings as well as a failure to safeguard users' personal data~\cite{PIPEDAEnforcement}.
South Korea's Personal Information Protection Commission found similar consent violations for Meta~\cite{SouthKoreaEnforcement}.
Under the CCPA the California Attorney General most recently brought a case against Disney for not properly opting out users, including via GPC, resulting in a \$2,750,000 settlement~\cite{CalAGCCPAEnforcementDisney}.
Similar major actions involved Healthline \cite{CalAGCCPAEnforcementHealthline} and Sephora~\cite{CalAGCCPAEnforcementSephora}.
Together with the Colorado and Connecticut Attorneys General the California Attorney General recently conducted a privacy enforcement sweep~\cite{CTCACOPrivacySweep}.
The new California Privacy Protection Agency (CPPA) also recently brought its first major enforcement actions against PlayOn Sports and Ford for not letting users opt out from tracking and adding unnecessary friction to the opt-out process, respectively~\cite{CPPACCPAEnforcementPlayOnSports, CPPACCPAEnforcementFord}.

\subsubsection{Low Enforcement Activity}

Brazil, India, Singapore, and South Africa are low enforcement jurisdictions.
These countries are either at the beginning of their privacy lawmaking and enforcement activities (Brazil and India) or have not engaged much in enforcement despite their privacy laws being effective for more than a decade (Singapore and South Africa).
Brazil's LGPD is closely modeled after the GDPR. 
However, the National Data Protection Authority is still in the early stages of its enforcement activities. 
While it has started issuing sanctions, those focus on foundational issues, such as responding to data breaches, rather than violations of user consent choices on websites~\cite{BrazilEnforcement}.
It is noteworthy that the first violations came, for the most part, from the public sector~\cite{BrazilEnforcement}.
India's DPDP Act, 2023 is even more recent than that of Brazil and enforcement has yet to begin. 
Both Singapore's PDPA and South Africa's POPIA date back more than a decade and have not been strongly enforced as to users' privacy rights and consent choices.

\section{Methodology}
\label{Methodology}


We conducted a country-localized web crawl using ten identically provisioned VM servers, one per country, to visit for each country the Common Top~525 and Country-specific Top~525 sites, derived from the Tranco list~\cite{LePochat2019} (We used the Tranco list version of  27 November 2023, available at \href{https://tranco-list.eu/list/Q9Z84}{https://tranco-list.eu/list/Q9Z84}).

\subsection{Common and Country-Specific Top 525 Sites}

\begin{enumerate}[topsep=3pt,parsep=0pt,itemsep=0pt,leftmargin=*,labelsep=5mm,align=parleft]
  \item Common Top 525 Sites:  
        This list contains the 525 most popular sites taken directly from the Tranco list. All 525 sites have a \texttt{.com} top level domain.
  \item Country‑specific Top 525 Sites:  
        To identify the most popular sites in a country we used sites' country code top-level domains.
        For example, we classified a site with \texttt{.de} domain as a German site.
\end{enumerate}

We chose not to create a separate Country-specific Top 525 list for the US because a substantial portion of the Common Top 525 sites are already US-based.
In addition, most US sites commonly use the \texttt{.com} top-level domain instead of the country-specific \texttt{.us} top-level domain, which, thus, would not accurately reflect the most popular sites in the US.
When preparing the crawl list for each country we started top-down in the Tranco list and manually loaded each site to ensure its availability or redirection to a loading site.
If a site failed to load, the browser returned an error page, or required human verification, we did not include the site and evaluated the next one until we arrived at 525 sites for a list.
We chose this number of sites to crawl to ensure a statistically meaningful sample for each country given that some sites' analyses may fail.
In total, we prepared nine Country‑specific Top 525 site lists and one Common Top 525 site list that doubles as Country-specific Top 525 list for the US.
Thus, our study covers $9 \times 1,050 + 525 = 9975$ sites across ten countries, corresponding to $525 \times 9 + 525 = 5250$ unique sites.

\subsection{VM Server Locations}

To ensure that the privacy laws of the ten countries we examine applies, we performed our crawls via the Google Cloud Platform on ten different VM servers located in the respective countries. 
Each VM was identically configured with 4 vCPUs, 16 GB of memory, 50 GB of storage, and the Windows operating system to ensure consistent performance across countries. 
Table~\ref{tab:VM-city-locations} shows the VM servers' geographic locations.

\vspace{-12pt}
\begin{table}[H]
    \centering
    \small
    \caption{List of VM servers and their geographic locations.}
          \newcolumntype{C}{>{\centering\arraybackslash}X} 
    \begin{tabularx}{\textwidth}{CCC}
    \toprule
    \textbf{Country} & \textbf{Continent} & \textbf{City} \\
    \midrule
    Australia       & Oceania        & Sydney \\
    Brazil          & South America  & São Paulo \\
    Canada          & North America  & Toronto \\
    Germany         & Europe         & Berlin \\
    India           & Asia           & New Delhi \\
    Singapore       & Asia           & Singapore \\
    South Africa    & Africa         & Johannesburg \\
    South Korea     & Asia           & Seoul \\
    Spain           & Europe         & Madrid \\
    US (California) & North America  & Los Angeles \\
    \bottomrule
    \end{tabularx}
    \label{tab:VM-city-locations}
\end{table}
\vspace{-16pt}

\subsection{Setup and Crawl Overview}

\begin{figure}[H]
    \centering
    \includegraphics[width=0.63\linewidth]{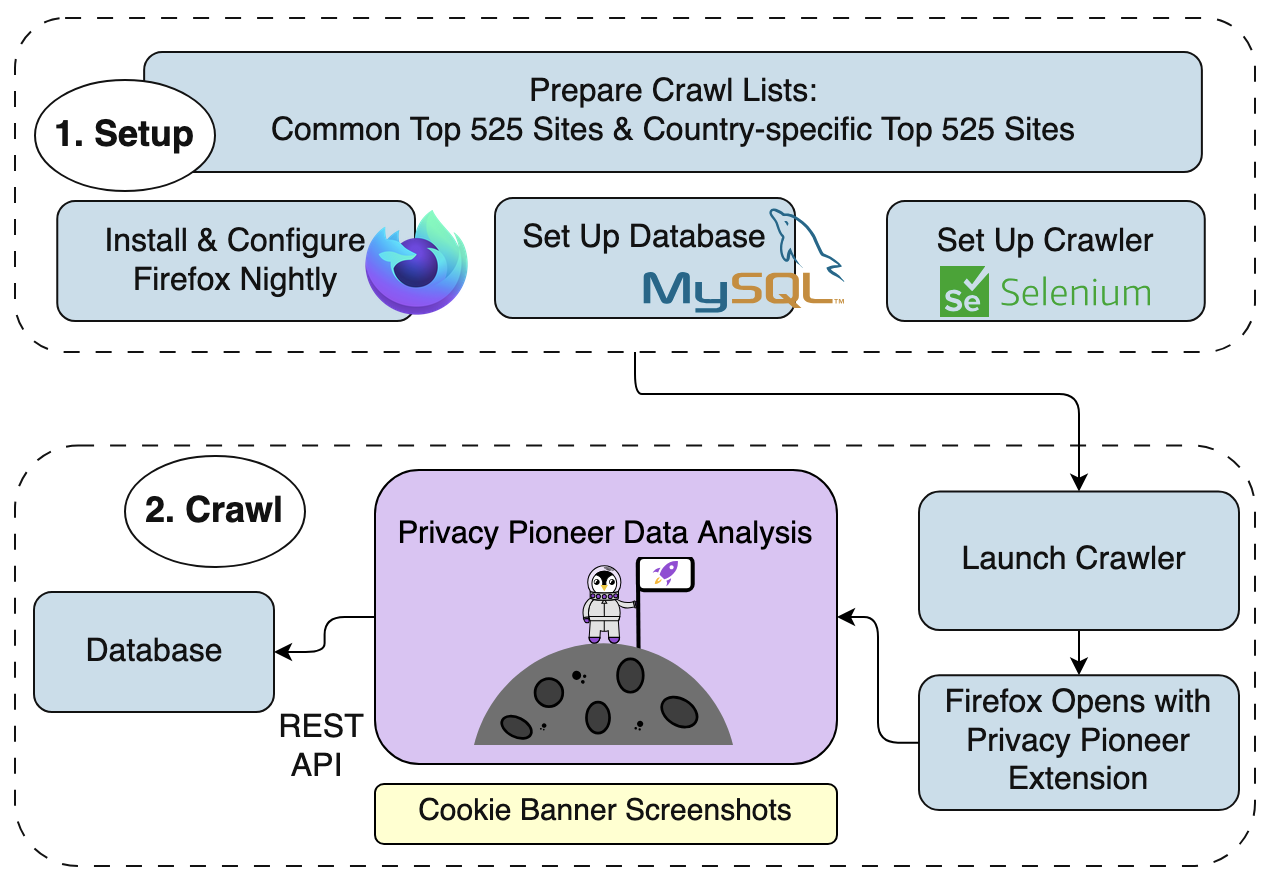} 
    \caption{Overview of our web crawling pipeline implemented and run on a separate VM server for each country.}
    \label{fig:privacy-pioneer-pipeline}
\end{figure}

We deployed our crawler with a Selenium WebDriver~\cite{selenium_webdriver} on Firefox Nightly with Enhanced Tracking Protection disabled~\cite{mozilla_etp_2024}. 
The crawler then installed the Privacy Pioneer browser extension~\cite{zimmeckEtAlPrivacyPioneer2024}, observed each site for 60 seconds (Figure~\ref{fig:Crawl_Timespan}), and persisted web requests and responses via a REST API to a local MySQL database~\cite{mysql_installer_download}. 
Based on the Disconnect Tracker Protection lists~\cite{disconnect_tracker_protection}, Privacy Pioneer labeled third-party connections into advertising, analytics, and social categories by matching request URLs against the URLs on the lists.
Figure~\ref{fig:privacy-pioneer-pipeline} illustrates our automated web crawling pipeline.

\vspace{-20pt}


\subsection{Entries}

\begin{figure}[H]
    \centering
    \includegraphics[width=3.9in]{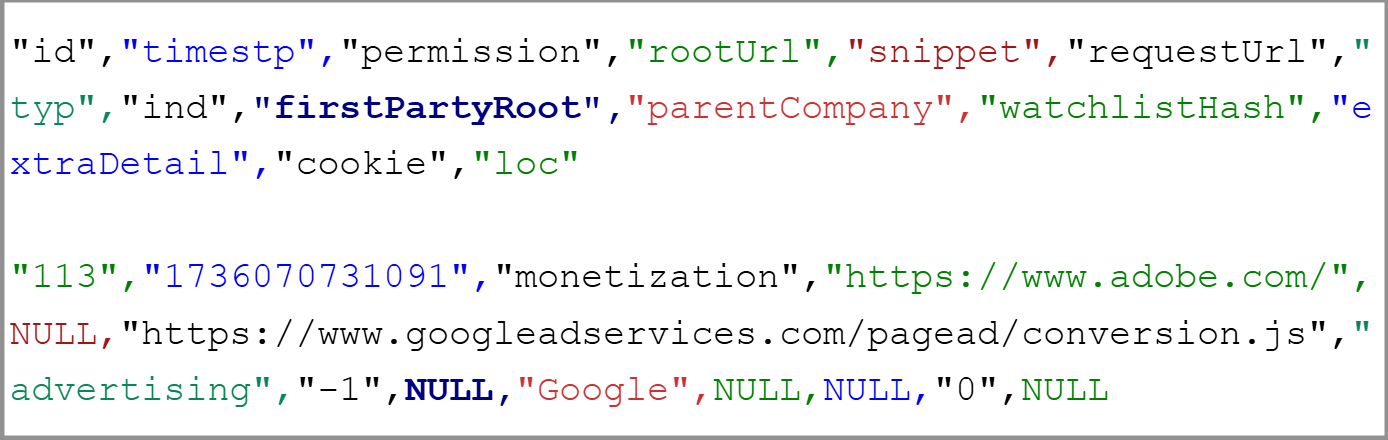}
    \caption{An example entry recorded by Privacy Pioneer. Upon visiting \href{https://www.adobe.com/}{https://www.adobe.com/} (``rootUrl''), the site loaded Google's conversion script (``requestUrl'' = \href{https://www.googleadservices.com/pagead/conversion.js}{https://www.googleadservices.com/pagead/conversion.js}). The entry is an advertising entry, i.e., labeled ``typ'' = \texttt{advertising}, ``parentCompany'' = \texttt{Google}, and ``cookie'' = \texttt{0} (no cookie recorded).}
    \label{fig:entry_example}
\end{figure}

Throughout this study, we use the term ``entry'' to refer to a unique record of an HTTP request for a tracking purpose, e.g., advertising, recorded by Privacy Pioneer. 
We refer to a ``tracker connection'' as the action that produces an entry. 
We use the number of entries as a proxy to quantify potential data sharing for a particular purpose with third~parties.
Figure~\ref{fig:entry_example} shows an advertising entry. 

\vspace{-10pt}

For clarity, if a site makes multiple requests to the same domain (i.e., the same top- and second-level domain) for the same purpose (e.g., for analytics to \href{https://google-analytics.com}{https://google-analytics.com}), we count it as a single entry. 
Connections to different domains (i.e., a different top- or second-level domain) are counted as distinct entries even if they are for the same purpose and belong to the same company (e.g., \href{https://connect.facebook.net}{https://connect.facebook.net} and \href{https://www.facebook.com}{https://www.facebook.com}).
However, different subdomains of the same top- and second-level domain (e.g., \href{https://ads.google.com}{https://ads.google.com} and \href{https://analytics.google.com}{https://analytics.google.com}) are counted as one entry.
A single HTTP request can contain multiple entries.

\subsection{Accuracy Test Results}

To ensure that it produces accurate results we tested our pipeline in its entirety---VM environment, crawler, and Privacy Pioneer extension---under conditions mimicking a real crawl. 
Table ~\ref{tab:ground-truth-analysis} shows the accuracy test results, which indicate reliable performance.

\begin{table}[H]
    \centering
    \small
        \caption{Accuracy test results for $n = 10$ sites. (TP = True Positive, FN = False Negative, and FP = False Positive).}
    \newcolumntype{C}{>{\centering\arraybackslash}X} 
    \begin{tabularx}{\textwidth}{CCCCCCC}
    \toprule
    \textbf{Category} & \textbf{TP} & \textbf{FN} & \textbf{FP} & \textbf{Precision} & \textbf{Recall} & \textbf{F1} \\
    \midrule
    Analytics                                & 43 & 3  & 0 & 1.00 & 0.93 & 0.97 \\
    Advertising                              & 126 & 0  & 0 & 1.00 & 1.00 & 1.00 \\
    Social                                   & 17 & 0  & 0 & 1.00 & 1.00 & 1.00 \\
    \bottomrule
    \end{tabularx}
    \label{tab:ground-truth-analysis}
\end{table}

We randomly selected ten sites from a subset of 100 sites for our test (The 100 sites contain five randomly selected sites from each Country-specific Top 525 list and 50 sites with a high likelihood of performing browser fingerprinting and other privacy-invasive practices that we identified via BuiltWith~\cite{builtwith}. 
Thus, the list is intended to identify practices beyond this study. 
For purposes of this study the sampling strategy of the 100 sites is less relevant as we focus here on the sites' integration of advertising, analytics, and social functionalities, which are common on many sites).
We performed our test on the US VM.
Given the uniformity of our setup across all countries, we have no reason to believe that performance would differ across countries. 
For each site, we downloaded the HTTP Archive (HAR) files containing complete records of all network requests and responses during the test crawl. 
These HAR files served as the basis for establishing the ground truth that we manually evaluated against the entries Privacy Pioneer flagged during the accuracy test run (Further details of our methodology---including its limitations---are described in \ref{Appendix:Methodology}).

\section{Results}
\label{Results}

\begin{enumerate}[label=\textbf{RQ\arabic*.},topsep=3pt,parsep=0pt,itemsep=0pt,leftmargin=*,labelsep=0.5mm,align=parleft] 
    \item Users' Geographic Location Is a Key Factor for Tracker Exposure Level
: Given the identical Common Top 525 sites for each country, we find different levels of tracker exposure.
    For the US (California) and Australia---two of the three opt-out jurisdictions in our study---sites have the highest level with 11.7 and 11.2 average entries per site, respectively, while opt-in jurisdictions generally have lower levels, especially, Spain and Germany with 5.3 and 4.2 average entries per site, respectively (Section~\ref{sec:country_level_data}
). 
    \item Globally Popular Sites Have a More Elevated and Narrower Tracker Range than Locally Popular Sites: Notably, only 44.6\% of German Country-specific Top 525 sites establish any tracker connection at all, while at the high end 96.0\% of Australian sites do so (Figure~\ref{fig:common_prop_datacollection_monetization}).
    The number of average entries per site for the Country-specific Top 525 sites ranges from 0.9 to 14.1, while the range for the Common Top 525 sites is narrower and overall more elevated with 4.2 to 11.7 average entries per site (Section~\ref{sec:country_level_data}).
    \item EU Law Appears to Reduce Tracking in EU Countries but Not as Much in Other Countries: 
    The Common Top 525 sites have 50.5\% fewer average tracker connections when accessed from EU countries compared to non-EU countries (Section~\ref{sec:country_level_data}), suggesting that the GDPR and ePrivacy Directive have a tangible effect in reducing tracking.
    Further, simply not interacting with cookie banners decreases tracker connections by 48.5\% for Germany, as measured for a sample of 36 Common Top 525 sites (Section~\ref{sec:Impact of Cookie Banner Consent and Inactivity}
).
    As 28\% of Common Top 525 sites show cookie banners in all ten countries (Section~\ref{sec:cookie_banner_accross_countries}
), our results suggest a moderate Brussels effect.
    However, against the backdrop of global US ad tech practices, EU law mainly acts as a Brussels shield.
\end{enumerate}

\subsection{Tracker Connections}

Overall, we successfully crawled 95.1\% (9488/9975) sites across ten countries.
The success rates of the Common and Country-specific Top 525 sites are similar.
For the Country-specific Top 525 sites (including the Common Top 525 sites for the US counted as Country-specific Top 525 sites) we successfully crawled 94.9\% (4984/5250) unique sites.
Unless mentioned otherwise, to ensure comparability for our analysis of the Common Top 525 sites we will use  only the 80\% (420/525) sites that we crawled successfully in all ten countries.

\subsubsection{Overall Tracker Connections and Background}

Table~\ref{tab:permission-entry-level} shows the percentages of entries identified and recorded across different categories. 
Table~\ref{tab:permission-row-level} shows the percentages of sites that connect to a third party in a category at least once.
While Table~\ref{tab:permission-entry-level} reflects the overall volume of tracker connections that sites in our dataset make to trackers in the various categories, Table~\ref{tab:permission-row-level} reflects their presence on the sites.
With 69.4\% of sites establishing an analytics connection, analytics is the most prevalent tracker category.
However, comparing 55,079 connections to advertising trackers to 19,760 analytics connections, the volume of the former is approximately 2.8 times of the latter.
This difference may be the result of underlying functional and economic reasons.

\begin{table}[H]
  \small
  \centering
        \caption{Entry-level counts ($n=84,170$) for all successfully crawled Country-specific and Common Top 525 sites ($n=9488$).}
    \newcolumntype{C}{>{\centering\arraybackslash}X} 
    \begin{tabularx}{\textwidth}{CCC}
      \toprule
      \textbf{Category} & \textbf{Entries Count} & \textbf{\% of Entries} \\
      \midrule
      Advertising & 55,079 & 65.4\% \\
      Analytics & 19,760 & 23.5\% \\
      Social & 9331 & 11.1\% \\
      \bottomrule
    \end{tabularx}
    \label{tab:permission-entry-level}
\end{table}
\vspace{-30pt}
  
\begin{table}[H]
    \centering
        \caption{Site-level counts ($n=8347$) for successfully crawled Country-specific Top 525 sites, with Common Top 525 sites for the US ($n=4984$).}
     \newcolumntype{C}{>{\centering\arraybackslash}X} 
    \begin{tabularx}{\textwidth}{CCC}
      \toprule
      \textbf{Category} & \textbf{Sites Count} & \textbf{\% of Sites} \\
      \midrule
      Advertising & 3060 & 61.4\% \\
      Analytics & 3460 & 69.4\% \\
      Social & 1827 & 36.7\% \\
      \bottomrule
    \end{tabularx}
    \label{tab:permission-row-level}
\end{table}
\vspace{-6pt}

Advertising trackers generally need to engage in more fine-grained tracking with more frequent connections as operators try to learn as much as possible about a user to serve relevant ads.
Analytics trackers, on the other hand, may only request a user agent string once and otherwise may not require frequent connections.
There is also an incentive for site operators to integrate multiple ad networks to maximize ad revenue while multiple analytics services would quickly lead to a duplication of functionality.
While social connections have the lowest level in both measurements, the industry is much more concentrated with only four parent companies---Facebook, LinkedIn, X, and reddit---in the top 30 trackers while both the advertising and analytics categories have a long tail (Section~\ref{sec:Impact of Cookie Banner Consent and Inactivity}
).
Overall, given their prevalence and volume, advertising trackers create the highest level of privacy risks for users.

\subsubsection{Percentage of Sites with Tracker Connections} 
\label{Percentage-of-Sites-with-Tracker-Connections}

\begin{figure}[H]
    \centering
    \includegraphics[width=4.2in]{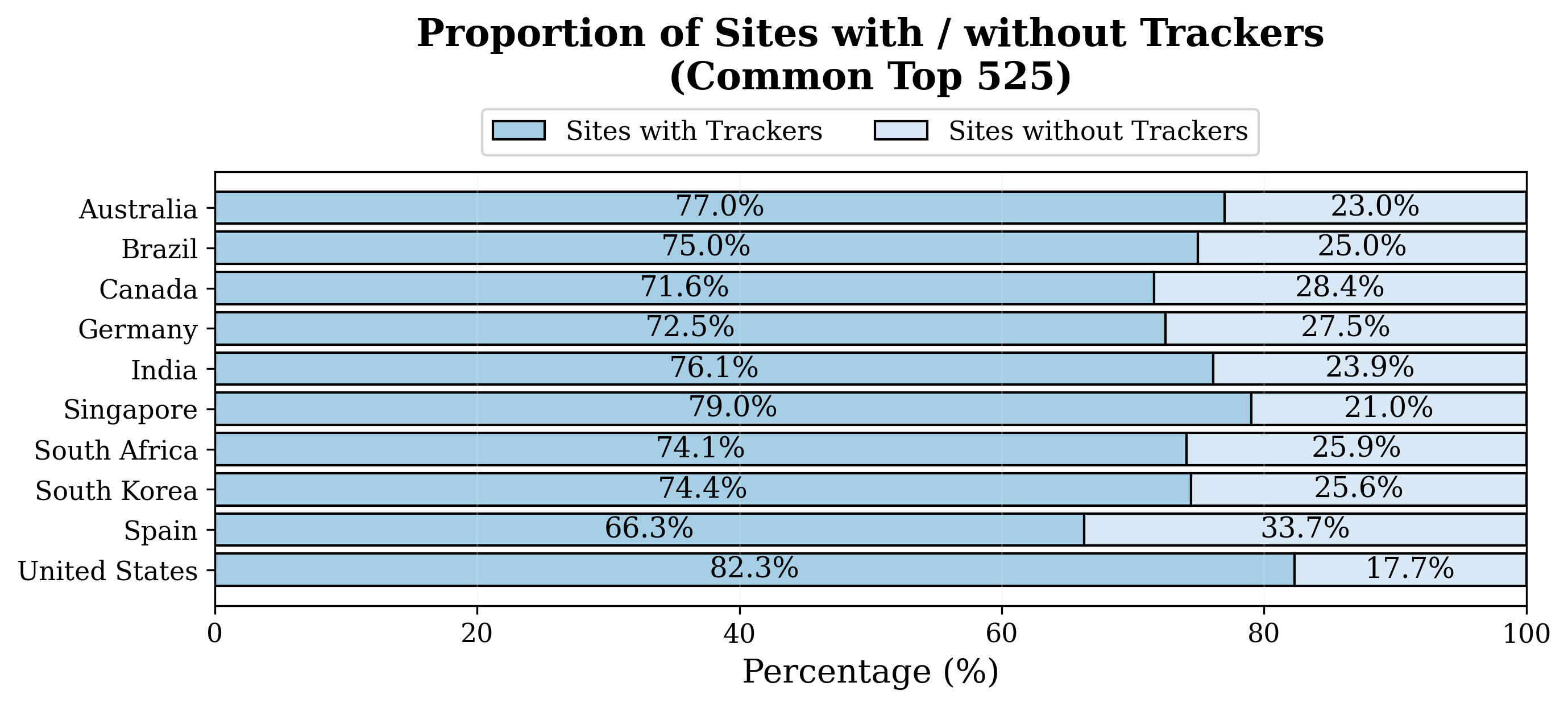}
    \includegraphics[width=4.2in]{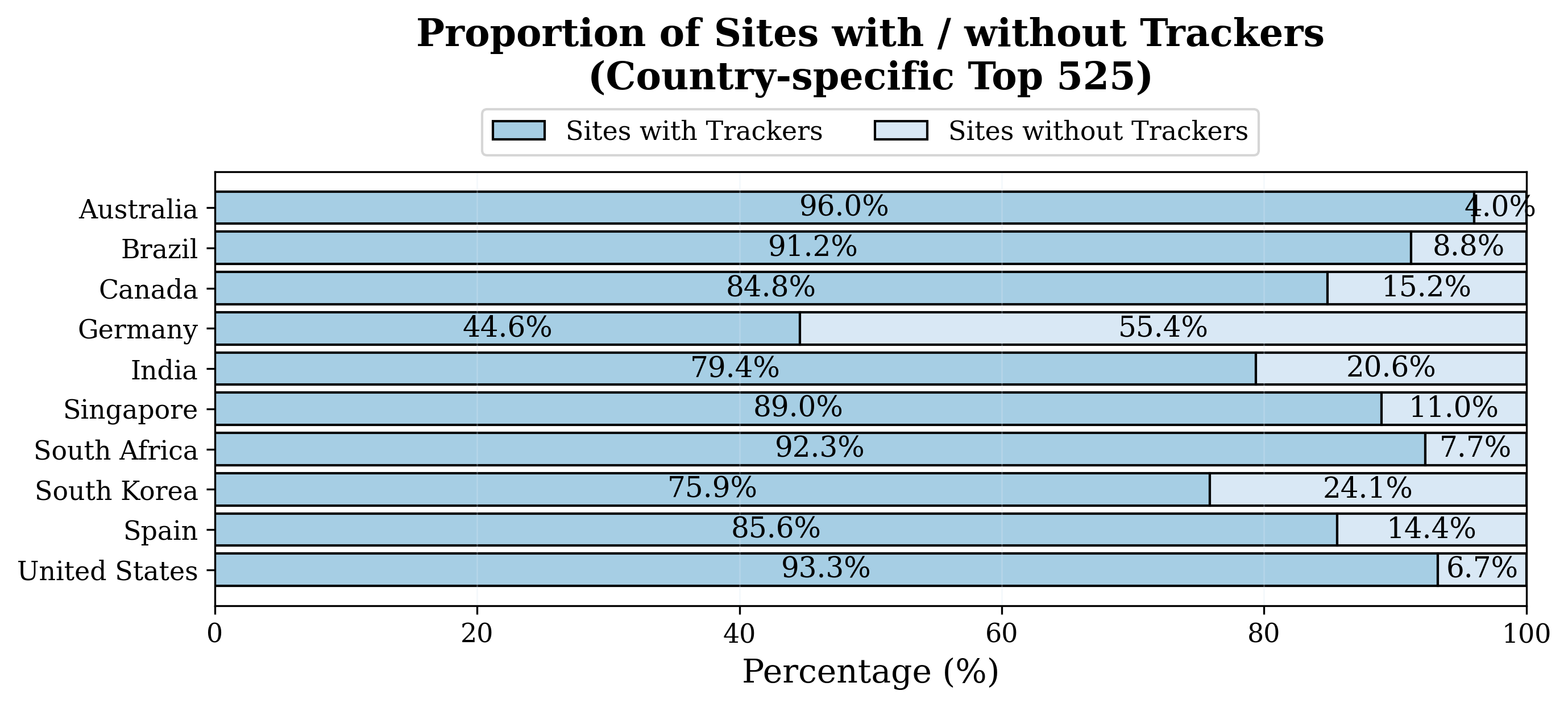}
    \caption{For the Common Top 525 sites (\textbf{top}) the percentages of sites with tracker connections across countries have a much smaller range compared to the Country-specific Top 525 sites (\textbf{bottom}). To ensure that data from different countries are comparable in the Common Top 525 sites we only include sites that we crawled successfully in all countries ($n = 420$ for each country), which is also the reason for the different percentages for the US in the Common and Country-specific Top 525 sites.}
    \label{fig:common_prop_datacollection_monetization}
\end{figure}

Figure~\ref{fig:common_prop_datacollection_monetization} shows the percentage of sites in a country that make at least one advertising, analytics, or social tracker connection. 

\vspace{-10pt}

Across all countries, between 17.7\% and 33.7\% of the Common Top 525 sites do not connect to any tracker. 
However, for the Country-specific Top 525 sites this range is 4.0\% to 55.4\%.
Compared to the other countries, Germany has a significantly higher percentage of sites that make no tracker connection. 
In contrast, Australia, Brazil, South Africa, and the US each have fewer than $10\%$ of sites not making such connections. 
This disparity suggests that the potential privacy risks associated with tracker connections differ between global and local sites.
Depending on which types of sites users visit, they are exposed to quite different levels of tracking.

\subsubsection{Average Number of Entries per Site} 
\label{sec:country_level_data}

\begin{figure}[H]
    \centering
    \includegraphics[width=3.6in]{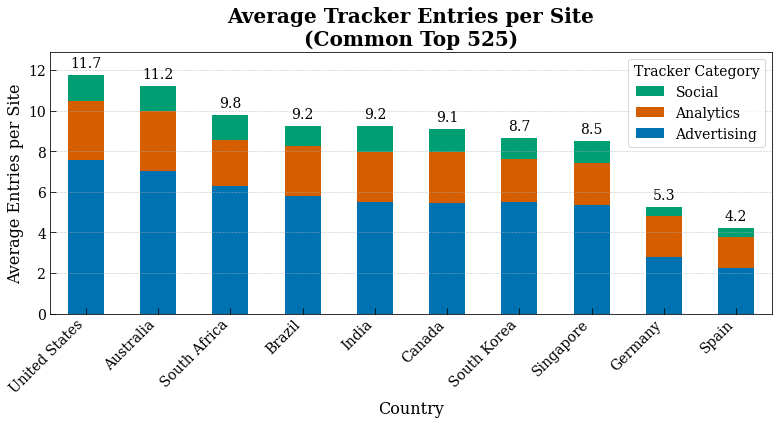}
    \includegraphics[width=3.6in]{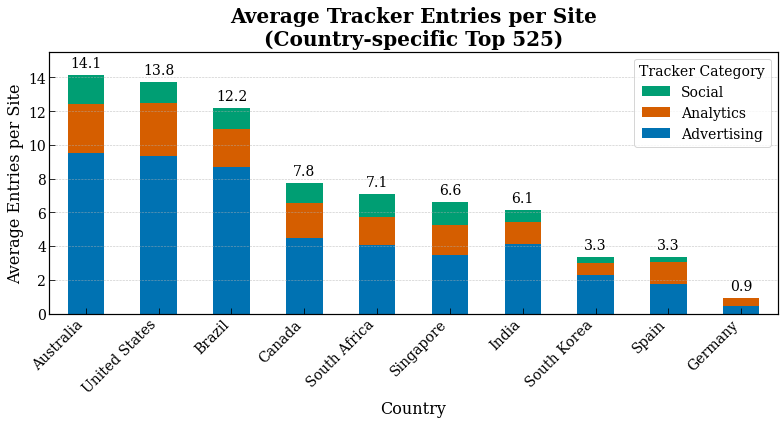}
    \caption{For both the Common Top 525 sites (\textbf{top}) and the Country-specific Top 525 sites (\textbf{bottom}), sites in the EU---Germany and Spain---have the lowest number of average entries. Australia and the US have the highest. (Common Top 525 sites: $n = 420$ for each country, Country-specific Top 525 sites: $n = 4984$ for all countries.)}
    \label{fig:average-entries-common525}
\end{figure}

Figure~\ref{fig:average-entries-common525} shows the average number of entries across countries. 
For most countries (South Africa, India, Canada, South Korea, Singapore, Germany, and Spain) it is greater in the Common Top 525 sites than in the Country-specific Top 525 sites.
The Country-specific Top 525 sites show a greater range from 0.9 to 14.1 compared to the Common Top 525 sites, which have a range from 4.2 to 11.7.
In the Country-specific Top 525 sites Australia exhibits over 14 times more average entries than Germany. 
However, for all countries except one (Germany), the percentage of sites making at least one tracker connection is greater in the Country-specific Top 525 sites (Figure~\ref{fig:common_prop_datacollection_monetization}).
For the Common Top 525 sites, on average, EU countries have 50.5\% fewer entries (4.8) than non-EU countries (9.7). 


Despite the differences in entry averages between the Common and Country-specific 525 sites, the overall ranking of countries across these two distributions remains generally consistent.
However, there are a few notable individual differences. South Korea's Country-specific Top 525 sites exhibit far fewer average entries than their Common Top 525 counterparts placing it on par with Spain. 
Conversely, Brazil's country-specific sites have more average entries than those of Canada, South Africa, Singapore, India, and South Korea---countries that are comparable to Brazil for in the Common Top 525 sites.
Overall, the following pattern emerges for the number of average entries per site across the Common and Country-specific Top 525 sites:

\begin{itemize}[topsep=3pt,parsep=0pt,itemsep=0pt,leftmargin=*,labelsep=5.8mm,align=parleft]
    \item High: Australia, US (California). 

    \item Medium: Brazil, Canada, India, Singapore, South Africa, South Korea.
    \item Low: Germany, Spain.
\end{itemize}

\subsection{Cookie Banner Deployment}

We evaluate the deployment of cookie banners on Common and Country-specific Top 525 sites. 
Independently of whether a jurisdiction requires opt-in consent or gives users a right to opt-out, generally all laws require notice.

\subsubsection{Tracker Connections by Sites with and Without Cookie Banners}

\begin{figure}[H]
    \centering
    \includegraphics[width=3.6in]{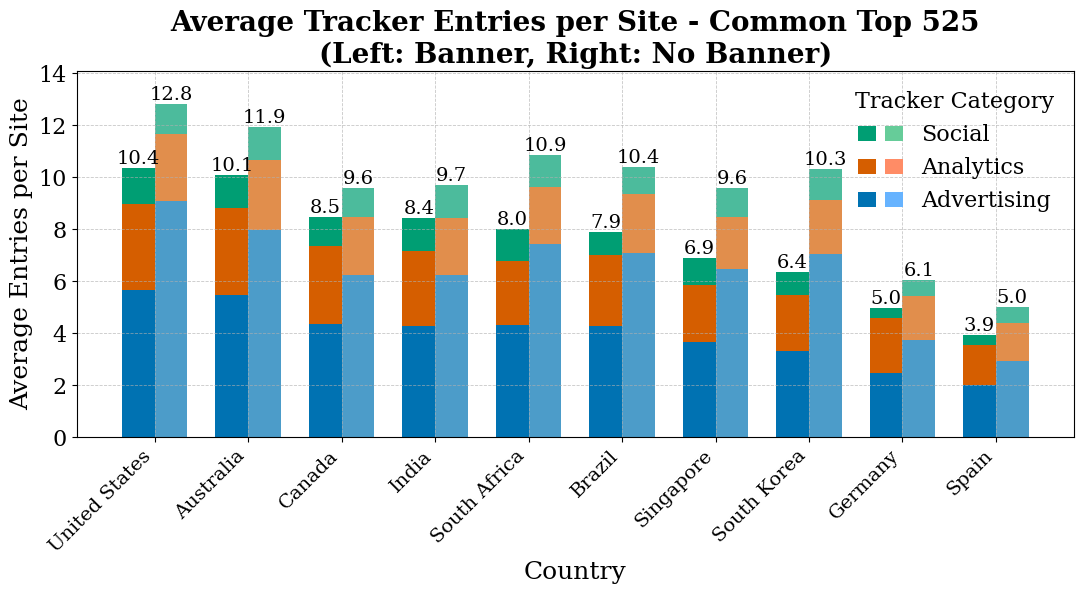}
    \includegraphics[width=3.6in]{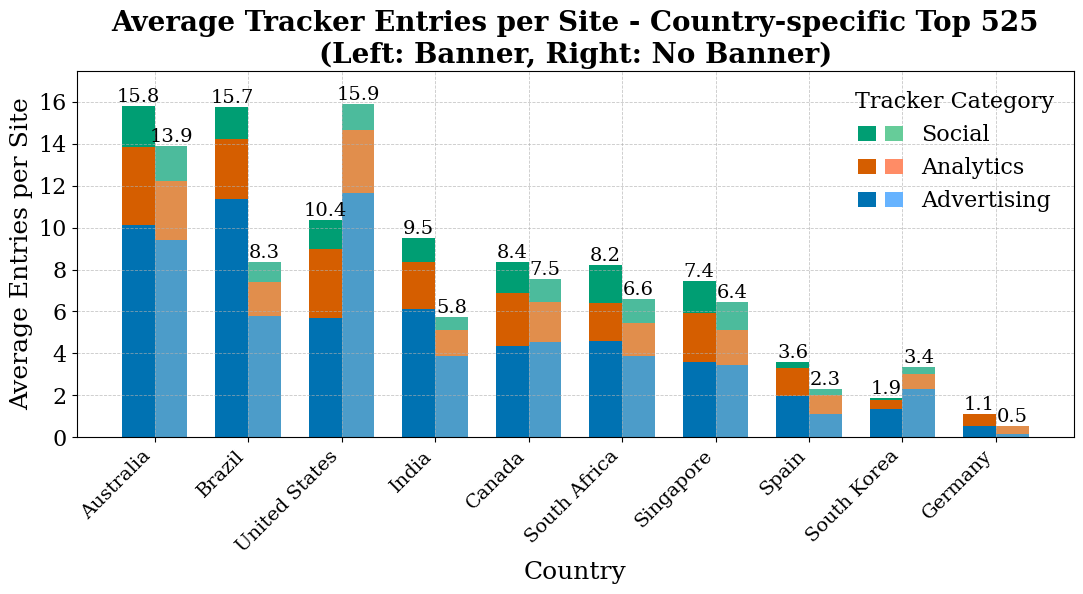}
    \caption {The Common Top 525 sites (\textbf{top}) with a banner have fewer average entries than sites without. However, this trend is reversed for the Country-specific Top 525 sites (\textbf{bottom}) except for the US (California) and South Korea. (Common Top 525 sites: $n = 420$ for each country, Country-specific Top 525 sites: $n = 4,984$ for all countries.)}
    \label{fig:baner_vs_nonbanner}
\end{figure}

Figure~\ref{fig:baner_vs_nonbanner} shows the average number of entries in the advertising, analytics, and social categories for the Common and Country-specific Top 525 sites split into sites with and without a cookie banner.
For the Common Top 525 sites, those without a banner have more average entries than sites with a banner.
This trend holds for all countries.
Sites without a banner have on average 9.6 entries, which is 26.3\% more than the 7.6 average entries we observe for sites with a banner. 
This difference is primarily due to more advertising entries. 
While Germany and Spain have a comparatively lower level of average entries, the trend of higher percentages for sites without a banner holds for them as well.

This finding suggests higher levels of non-compliance for sites without a cookie banner.
It is expected that sites without a banner would not connect to third parties (or only connect to third parties that do not require consent). 
However, our finding suggest the opposite: sites without a banner are more likely to violate applicable privacy laws as they generally connect to advertising trackers at higher rates.
From a user's perspective, not seeing a banner is less disruptive but also more privacy-invasive as sites share the user's personal data ``silently''.

Comparing banner and non-banner sites for the Country-specific Top 525 sites, we see the opposite trend compared to the Common Top 525 sites: non-banner sites generally have fewer average entries than banner sites indicating less broad privacy risk. 
In particular, we observe a substantial decrease in average entries for non-banner sites for Germany (54.5\%), Brazil (47.1\%), India (38.9\%), and Spain (36.1\%).
However, while the average number of entries for non-banner sites is lower, there still can be a high level of non-compliance as the banner requirement applies regardless of how many tracker connections a site makes as long it is at least one.



\subsubsection{Cookie Banner Deployment Across Countries}
\label{Cookie-Banner-Distribution-Across-Countries}


 
\begin{figure}[H]
    \centering
    \includegraphics[width=3.4in]{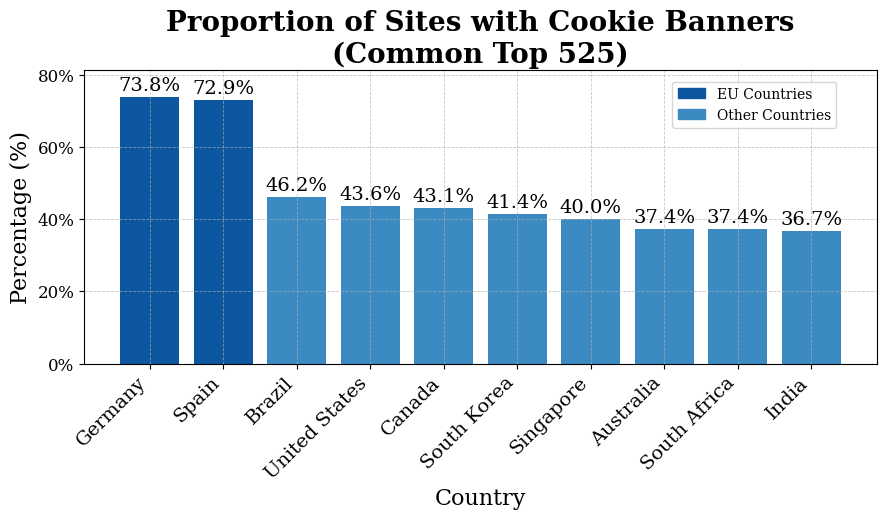}
    \includegraphics[width=3.4in]{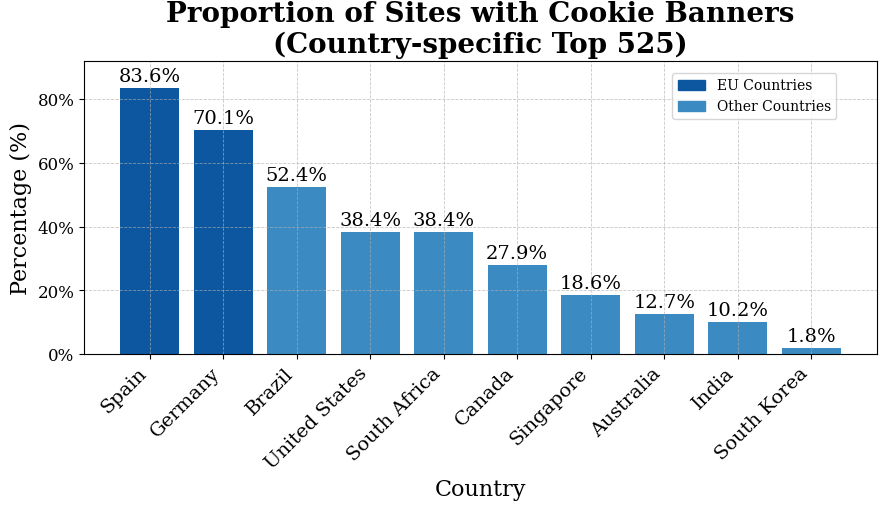}
    \caption{For the Common Top 525 sites the percentage of sites with cookie banners is in a narrow range across two sets of sites with one set consisting of the EU countries, Germany and Spain, and the other of the remaining countries (\textbf{top}). The banner distribution for the Country-specific Top 525 sites has a much larger range (\textbf{bottom}). (Common Top 525 sites: $n = 420$ for each country, Country-specific Top 525 sites: $n = 4984$ for all countries.)}
    \label{fig:cookie-banner-counts}
\end{figure}

For the Common Top 525 sites we observe substantially different distributions of cookie banner deployment between EU and non-EU countries.
Germany and Spain have the highest percentage of sites with cookie banners with 73.8\% and 72.9\% of sites, respectively. 
In contrast, the sites in the remaining eight non-EU countries have substantially lower percentages of cookie banner deployment ranging narrowly from 36.7\% to 46.2\%.
Figure~\ref{fig:cookie-banner-counts} shows the proportion of cookie banner deployment for both the Common and Country-specific Top 525 sites.

\vspace{-10pt}

For the Country-specific Top 525 sites, we see a much larger range across countries. 
For half of the countries the ranking by cookie banner deployment remains comparable to the Common Top 525 sites. 
Spain and Germany continue to exhibit the highest banner percentages. 
Brazil and the US (California) continue to rank third and fourth, respectively. 
South Africa shows only a minimal percentage difference between the two distributions.
However, for Canada, Singapore, Australia, India, and, in particular, South Korea, we observe substantially lower rates of banner deployment. 
On average, the percentage of sites with banners across these five countries drops from 39.7\% to only 14.2\%. 
South Korea represents the most extreme case with only 1.8\% of sites with cookie banners. 
These significantly lower levels of banners could indicate lower levels of privacy law compliance.

For South Korea, for example, 75.9\% of the Country-specific Top 525 sites connect to at least one tracker (Figure~\ref{fig:common_prop_datacollection_monetization}).
At the same time, only 1.8\% of sites deploy a cookie banner.
Thus, even though South Korea's privacy law mirrors the GDPR's requirement for opt-in consent, there may be a substantial compliance gap.
The higher and approximately uniform rates of cookie banner deployment among the non-EU Common Top 525 sites could be based on the Brussels effect.
Some site operators of globally popular sites may be simplifying their compliance obligations by EU cookie requirements for all their sites.
We can gain further insight into the strategies of site operators' deployment of cookie banners across countries by examining the Common Top 525 sites in this regard.

\subsubsection{Sites' Multi-country Cookie Banner Strategies} 
\label{sec:cookie_banner_accross_countries}

\begin{figure}[H] 
    \centering 
    \includegraphics[width=3.7in]{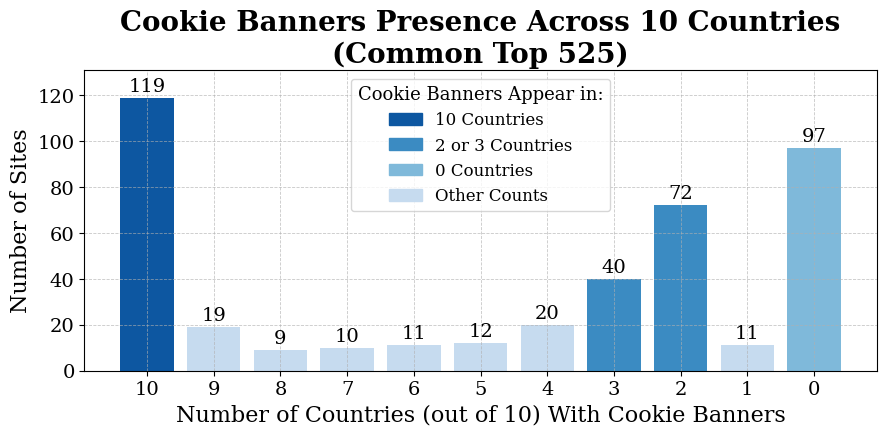} 
    \caption{Cookie banners across ten countries for the Common Top 525 sites ($n = 420$ for each country). 28\% (119/420) of sites show cookie banners in all ten countries. As previous findings suggest~\cite{10.1145/3313831.3376321,10.1145/3706598.3713648}, CMPs have substantial potential impact on cookie banner and tracker settings. Our findings point to their prominent role as well. During our crawl we observed 23.6\% (990/4200) connections to OneTrust, as indicated by a request to \href{https://geolocation.onetrust.com}{https://geolocation.onetrust.com}.} 
    \label{fig:cookie_banner_distribution} 
\end{figure}

Based on the directly comparable Common Top 525 sites, we can discern three major strategies: site operators deploy banners (1) everywhere, (2) nowhere, or (3) in two or three countries.
Figure~\ref{fig:cookie_banner_distribution} illustrates these strategies.
The x-axis values show the number of countries in which a site displays a cookie banner while the y-axis shows the count of such sites. 
For example, the leftmost bar shows that 119 sites display cookie banners in all ten countries. 
The rightmost bar, on the other hand, shows that 97 sites do not show a cookie banner in any country.
We observe a clear pattern for the majority of sites: if site operators choose to deploy a cookie banner, they will do so in either all countries or in two or three countries.

\vspace{-22pt}



Tables~\ref{tab:banner-2-countries} and~\ref{tab:banner-3-countries} show the sets of countries for which site operators deploy cookie banners in exactly two or three countries, respectively. 
Notably, almost all sites in the two-country set deploy banners only for their German and Spanish sites.
Site operators seem to put strong emphasis on compliance with the EU jurisdictions Spain and Germany.
This pattern suggests that compliance efforts in this regard are primarily driven by the GDPR and the ePrivacy Directive.
While comparable consent requirements exist in several non‑EU jurisdictions, including Brazil, South Korea, and, most recently, India, we do not yet observe any major impact by those.
However, for site operators that deploy cookie banners in exactly three countries, the set consisting of Brazil, Germany, and Spain is the second-most common (Table~\ref{tab:banner-3-countries}).
Increased enforcement of the LGPD could lead to further progress.
The same is true for opt-out notices under the CCPA that are part of the most common set of three countries with 16 sites.

\vspace{-16pt}
\begin{table}[H]
\centering
\small
    \centering
        \caption{Common Top 525 sites deploying banners in exactly two countries ($n = 72$ for each country).}
    \newcolumntype{C}{>{\centering\arraybackslash}X} 
    \begin{tabularx}{\textwidth}{CC}
    \toprule
    \textbf{Country Set} & \textbf{Sites} \\
    \midrule
    Germany and Spain             & 68 \\
    Australia and South Korea     & 1  \\
    Brazil and Germany            & 1  \\
    Germany and US (California)   & 1  \\
    Germany and India             & 1  \\
    \bottomrule
    \end{tabularx}
    \label{tab:banner-2-countries}
\end{table}
\vspace{-22pt}

\begin{table}[H]
    \centering
        \caption{Common Top 525 sites deploying banners in exactly three countries ($n = 40$ for each country).}
    \newcolumntype{C}{>{\centering\arraybackslash}X} 
    \begin{tabularx}{\textwidth}{CC}
    \toprule
    \textbf{Country Set} & \textbf{Sites} \\
    \midrule
    Germany, Spain and US (California)  & 16 \\
    Brazil, Germany and Spain           & 10 \\
    Germany, South Korea and Spain      & 5  \\
    Canada, Germany and Spain           & 4  \\
    Germany, South Africa and Spain     & 3  \\
    Australia, Germany and Spain        & 1  \\
    Singapore, South Africa and Spain   & 1  \\
    \bottomrule
    \end{tabularx}
    \label{tab:banner-3-countries}
\end{table}

\subsubsection{Impact of Cookie Banner Consent and Inactivity}
\label{sec:Impact of Cookie Banner Consent and Inactivity}





\begin{figure}[H]
    \centering
    \includegraphics[width=3.13in]{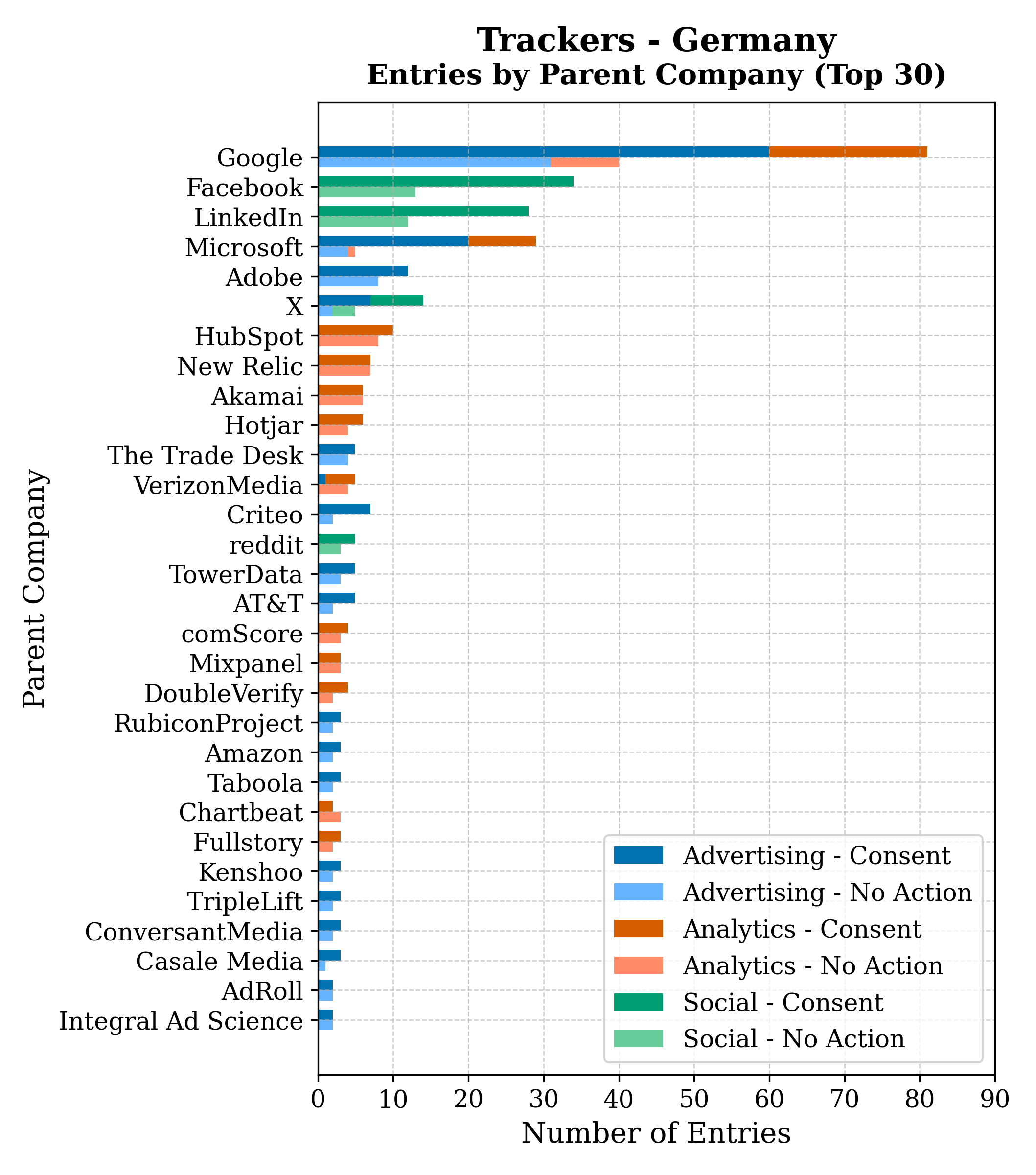}
    \includegraphics[width=2.9in]{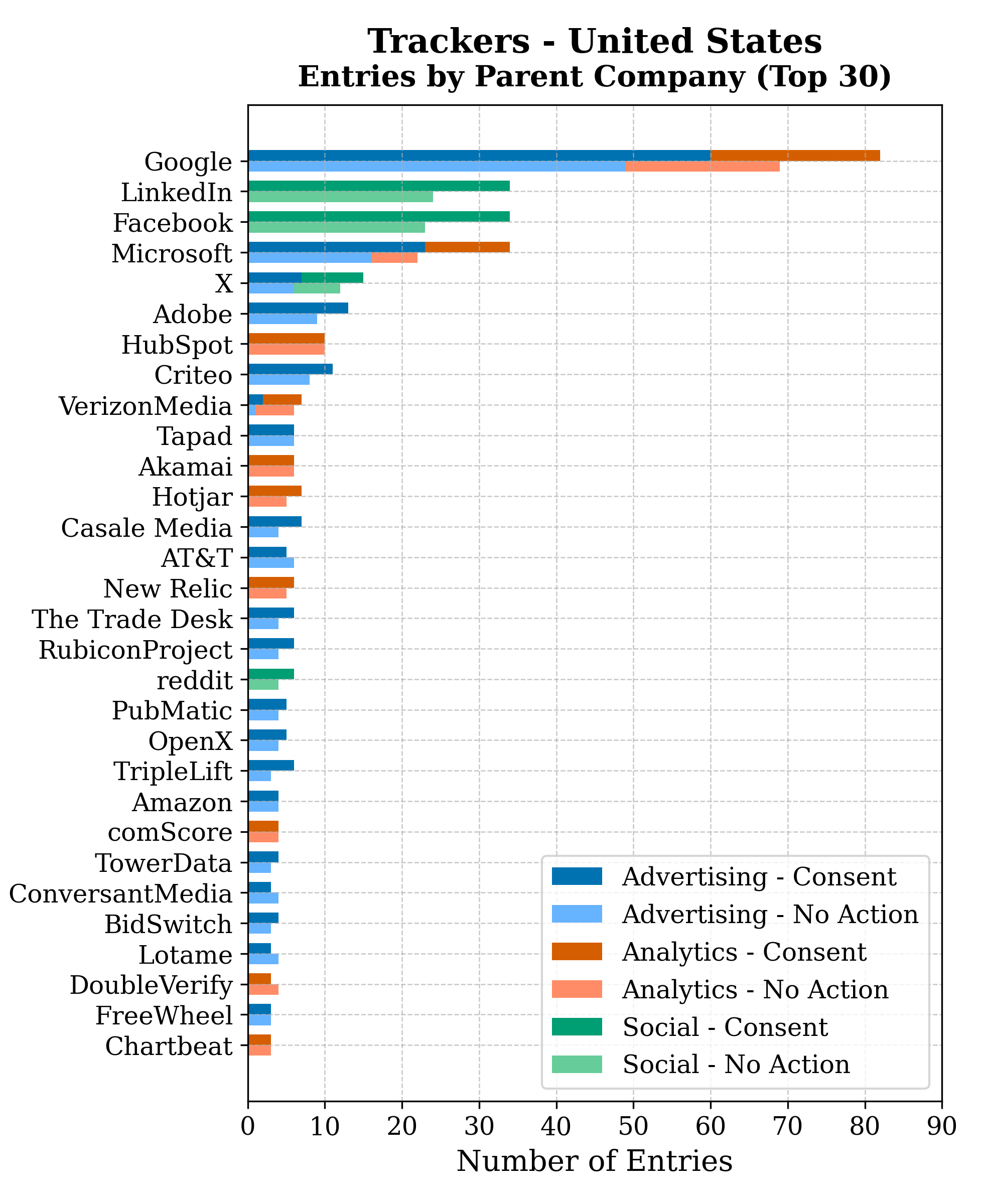}
    \caption{For Germany (left) the number of tracker connections upon not interacting with cookie banners is often reduced by at least half compared to consenting. The difference is substantially less pronounced for the US (right). Our measurement for Spain (Figure~\ref{fig:Sampled_50_Sites_1stPartyExcluded_errorsExcluded_Spain}) exhibits a similar trend as for Germany. (Common Top 525 sites: $n = 36$ for each country.) Figure~\ref{fig:parent-company-country-top-525} shows the distribution of parent companies for the Country-specific Top 525 countries.}     \label{fig:Sampled_50_Sites_1stPartyExcluded_errorsExcluded}
\end{figure}

Based on a random sample of 50 sites from the Common Top 525 sites for Germany, Spain, and the US, from which we exclude 14 sites that did not load properly in at least one country, we determine the impact of cookie consent and inactivity. 
Figure~\ref{fig:Sampled_50_Sites_1stPartyExcluded_errorsExcluded} shows the 30 parent companies with the highest entry counts for Germany and the US (See \ref{privacy-pioneer} for details on how we derive parent companies).
In particular, trackers in the advertising and social categories are substantially reduced (Table~\ref{tab:percentage_reduction}), while the rate of analytics trackers, apart from those of Google and Microsoft, shows a smaller reduction.
Overall, Microsoft's tracker rate was reduced the most.
As the set of top 30 parent companies is very similar across countries---with Google, Facebook, LinkedIn, Microsoft, Adobe, and X dominating the industry---any privacy improvements in their practices would have broad impact for many users.

\vspace{-30pt}

\begin{table}[H]
\centering
\caption{The percentage decrease of trackers when taking no action compared to consenting calculated for the data in Figure~\ref{fig:Sampled_50_Sites_1stPartyExcluded_errorsExcluded}. The overall decrease refers to advertising, analytics, and social categories combined.}
\small
\newcolumntype{C}{>{\centering\arraybackslash}X} 
\begin{tabularx}{\textwidth}{CCCCC}
\toprule
\textbf{Country} & \textbf{Advertising} & \textbf{Analytics} & \textbf{Social} & \textbf{Overall} \\
\midrule
Germany & $-$51.7\% & $-$34.5\% & $-$58.1\% & $-$48.5\%  \\
US (California) & $-$20.8\% & $-$11.7\% & $-$30.5\% & $-$21.1\% \\
\bottomrule
\end{tabularx}
\label{tab:percentage_reduction}
\end{table}

Comparing across countries, we observe much fewer tracker connections for Germany and Spain upon inactivity than for the US.
This observation aligns with the opt-in requirements of the GDPR and ePrivacy Directive.
If followed, both laws in combination create a twofold effect:
The ePrivacy Directive prevents sites from setting cookies as Article 5(3) gives users a right to refuse accessing or storing information on the user's terminal equipment.
The GDPR requires sites to obtain users' consent before initiating the network connection to trackers' servers, which would transmit users' IP addresses, that is, personal data under the GDPR.
As the CCPA permits tracking until users opt out, it is plausible that we observe a smaller tracker difference rate between consent and inactivity for the US (California).
Opt-in jurisdictions are inherently more privacy-protective.


%
%
%

\section{Discussion}
\label{Discussion}

Our results have broader implications for protecting web users' privacy and evolving the online ad ecosystem towards a more privacy-protective system that supports the free and open web for its users.

\subsection{Challenges for Users, Website Operators, and Lawmakers and Regulators}

Overall, our results show that geographic location is a key factor for the level of web tracking to which a user is exposed. 
This geographic fragmentation creates distinct challenges for three main groups: (1) users, (2) website operators, and (3) lawmakers and regulators.
Addressing these challenges in a coordinated and integrated approach is crucial for improving web privacy overall.

Currently, users experience different levels of privacy protections depending on applicable law, its enforcement, and implementation.
For example, only 44.6\% of German Country-specific Top 525 sites establish a tracker connection while 93.3\% of US sites do so (Figure~\ref{fig:common_prop_datacollection_monetization}). 
In general, EU users benefit from stronger legal protections but also face a higher frequency of cookie banners (Figure~\ref{fig:cookie-banner-counts}), which can create usability challenges and undermine the very purpose of meaningful consent. 
Users outside the EU---even in countries with strong privacy laws---often receive much less effective legal protection from tracking.
In many cases users are purposely misled via dark patterns~\cite{10.1145/3313831.3376321}.
Automating user rights, for example, via GPC~\cite{hausladenEtAlGPCWeb2025,zimmeckEtAlGPC2023}, which has been shown to be effective in reducing the number of intractable cookies~\cite{rasaii2025intractablecookiecrumbsunveiling}, can increase users' privacy protection while at the same time prevent the degradation of usability.

For website operators it can be a complex and costly endeavor to ensure privacy compliance with all applicable laws in the geographically fragmented legal landscape. 
They can implement robust controls for the EU while adopting a more permissive tracking approach elsewhere (Figure~\ref{fig:cookie_banner_distribution}). 
This bifurcated strategy, however, may become increasingly untenable as more countries adopt privacy laws with local nuances. 
The challenge is to develop a scalable compliance architecture for the web that respects local laws without creating an unmanageable patchwork of different rules for website operators.
That is the challenge that CMPs identified as their business case and that, for example, the Interactive Advertising Bureau is responding to with its Global Privacy Platform~\cite{IAB_GPP}.
These aggregation solutions can help operators' to make their sites compliant if they reflect the law adequately, especially, in their default settings, and are easy and cost-effective to implement.   

For lawmakers and regulators the challenge is to draft laws and regulations that are easy to implement and practical to enforce. 
Ideally, laws and regulations would converge across jurisdictions to reduce compliance overhead.
While such convergence may happen naturally on a practical level, it can also be explicit.
For example, for the right to opt out per the Texas Data Privacy and Security Act, operators are only required to support a technology if they use it for ``complying with similar or identical laws or regulations of another state'' (Texas Data Privacy and Security Act, Section 541.055(e)(4)).
This provision illustrates how lawmakers and regulators can support the convergence of privacy laws across different jurisdictions.
It also illustrates how laws and regulations can leave sufficient room for new privacy technology, for example, encompassing GPC.

\subsection{The Leading Role of the EU: Brussels Shield and Brussels Effect}

Our study serves as yet another example of the leading role the EU plays in shaping web privacy. 
Our results suggest that the GDPR and ePrivacy Directive are not merely laws on the books but have a tangible effect on reducing tracker connections. 
When accessed from Germany and Spain, on average, the Common Top 525 sites made 50.5\% fewer tracker connections compared to non-EU countries (Section~\ref{sec:country_level_data}). 
Also, simply not interacting with cookie banners decreased tracker connections by 48.5\% for Germany (Table~\ref{tab:percentage_reduction}).
These findings illustrate that under EU law privacy is protected by default.

The effect of EU law also appears to be reflected in the deployment of cookie banners. 
For the Common Top 525 sites we observed a trimodal distribution: sites tend to show banners in either (1) all countries, (2) primarily EU countries, or (3) no countries (Figure~\ref{fig:cookie_banner_distribution}). 
For the 28\% of Common Top 525 sites showing cookie banners in all countries our results suggest a moderate Brussels effect.
However, for many global sites (Figure~\ref{fig:cookie-banner-counts}) and third party operators (Figure~\ref{fig:Sampled_50_Sites_1stPartyExcluded_errorsExcluded}) the effect is localized to EU jurisdictions, meaning the law acts as a Brussels shield for EU users but does not broadly improve privacy for the rest of the world.
Thus, while the EU is playing a leading role in the adoption of consent mechanisms, its influence is also limited by geo-fencing.

\subsection{Enforcement as a Driver of Privacy Evolution}

Our results suggest that enacting a new privacy law by itself is insufficient to improve web privacy.
Rather, it is the combination of privacy lawmaking, regulatory rulemaking, and active enforcement that is most impactful. 
Various countries in our study are at different evolutionary stages of developing their privacy enforcement regime (Section~\ref{Privacy-Law-Enforcement}). 
Germany and Spain represent a mature stage with high regulatory awareness and active Data Protection Authorities and courts.
Brazil, on the other hand, is at an earlier stage.
While its LGPD mirrors the GDPR, it is still developing its enforcement capacity and has yet to build and apply an enforcement infrastructure that motivates and polices compliance.
As an illustration, we found the percentage of Common Top 525 sites with tracker connections for Brazil to be similar to the percentages for Germany and Spain (Figure~\ref{fig:common_prop_datacollection_monetization}).
However, the proportion of sites deploying cookie banners is substantially lower (Figure~\ref{fig:cookie-banner-counts}).
This discrepancy points to a gap between the law as written and its practical enforcement. 

The example of the EU shows that privacy reform is best accompanied by investment in the institutional and procedural infrastructure for providing guidance, monitoring compliance and, if necessary, acting on violations.
However, learning from other countries' efforts is not a one-way-street. 
For example, currently GPC has significant more adoption in the US compared to the EU.
California~\cite{XavierBecerraGPCTweet2}, Colorado~\cite{GPCColorado}, Connecticut~\cite{ConnecticutDataPrivacyActFAQs}, New Jersey~\cite{NewJerseyDataPrivacyLawFAQs}, and Oregon~\cite{OregonConsumerPrivacy} recognize GPC as a valid opt-out mechanism enforcing it accordingly~\cite{CalAGCCPAEnforcementDisney, CalAGCCPAEnforcementHealthline, CalAGCCPAEnforcementSephora, CTCACOPrivacySweep}. 
With its Digital Omnibus~\cite{DigitalOmnibus} the EU is now moving towards the adoption of privacy preference signals as well.
Adopting what works in other jurisdictions, in particular, when it comes to enforcement, does not only allow jurisdictions to profit from experiences others already made but also avoids fragmentation and contributes to the global convergence of privacy mechanisms and their enforcement.

\subsection{Maintaining a Free and Open Web}

Finally, we must acknowledge that a major driving force behind the web's privacy problem is the ad-financed online economy. 
Web tracking is the result of economic incentives in the online ad industry, which funds a large portion of the free content and services on the web available today and likely in the future.
Thus, in our view, the policy challenge is not to eliminate online ads altogether but to ensure that they are transparent, fair, and privacy-preserving. 
Evolving the web into an inherently privacy-protective system while maintaining the underlying economic benefits of its ad ecosystem is a delicate balancing act. 
Future efforts must ensure that the web remains free and open without requiring users to trade their privacy for access to content.

\section{Conclusions}
\label{Conclusions}

Across ten countries and 9,488 websites, the results of our study suggest that users' geographic location is a key factor for their tracker exposure level.
We find that the global Common Top 525 sites visited from EU jurisdictions---governed by the GDPR and ePrivacy Directive---exhibit 50.5\% fewer tracker connections than sites from non-EU jurisdictions. 
Viewed in light of the EU's regulatory activities, these findings suggest that privacy laws with active enforcement can meaningfully limit tracking, whereas laws that do not yet have regulatory follow-through, such as Brazil's LGPD, appear to be less effective.
While our results suggest a moderate Brussels effect for cookie banner deployment on global sites, EU law primarily acts as a Brussels shield against the backdrop of global US ad tech practices.

Looking ahead, we want to extend this study in various directions. 
First, longitudinal measurements can help gauging where future enforcement actions are necessary and where past enforcement actions have led to tangible privacy protections for users.
Second, we want to extend our measurements from manual consent mechanisms to privacy preference signals, such as GPC. 
Third, a closer integration of privacy laws into the technological structure of the web, for example, by means of web standards is desirable.
Overall, we want to evolve the web into an inherently privacy-preserving system that protects its users but also maintains and extends its free and open nature.


 
		\section*{Author Contributions}
H.Y.: data curation, investigation, methodology, software, validation, visualization, writing---original draft preparation; P.Y.: data curation, investigation, methodology, software, validation, visualization, writing---original draft preparation; S.Z.: conceptualization, investigation, methodology, software, validation, visualization, writing---original draft preparation, writing---reviewing and editing.
All authors have read and agreed to the published version of the manuscript.

		\section*{Funding}
This research was funded by Google under the Research Scholar Program (project title ``Privacy Pioneer: Automating the Creation of Privacy Labels for Websites'').
 
		\section*{Institutional Review Board Statement}
Not applicable.

		\section*{Informed Consent Statement}
Not applicable.

		\section*{Data Availability Statement}
        Data related to this study is publicly available under the MIT License~\cite{privacyPioneerWebCrawler}.
 
		\section*{Acknowledgments}
We would like to thank our anonymous reviewers for their detailed and valuable feedback that helped us to improve our paper substantially.
We would also like to thank Nina Taft and Hamza Harkous of Google's Applied Privacy Research group for the discussions and support of this project.
Brian Tang shared his expertise on cookie consent violations, for which we thank him kindly.
We would finally like to thank Wesleyan University, its Department of Mathematics and Computer Science, and the Anil
Fernando Endowment for their additional support. Conclusions reached or positions taken are our own and not necessarily those of our supporters, its trustees, officers, or staff.

		\section*{Conflicts of Interest}
The authors declare no conflict of interest. 

%
%

		\section*{Use of AI and AI-Assisted Technologies}

No AI tools were utilized for this paper.

\setcounter{section}{1}
\renewcommand{\thesection}{A.\arabic{section}} 


\appendix

%

\setcounter{table}{0}
\setcounter{figure}{0}
\renewcommand{\thesection}{Appendix \Alph{section}}
\renewcommand{\thetable}{A\arabic{table}}
\renewcommand{\thefigure}{A\arabic{figure}}

\section{Methodology Details}
\label{Appendix:Methodology}

\subsection{Setup and Crawl Details}
\label{sec:methodology-overview}

\begin{enumerate} [topsep=3pt,parsep=0pt,itemsep=0pt,leftmargin=*,labelsep=5mm,align=parleft]
    \item Crawl Environment Setup: For each country, we crawled sites from the Common Top 525 and Country-specific Top 525 sites. 
    We configured the required components on each country's VM, including the Firefox Nightly browser (We used the Firefox Nightly build of  1 January 2024, available at \href{https://ftp.mozilla.org/pub/firefox/nightly/2024/01/2024-01-01-23-15-40-mozilla-central/}{https://ftp.mozilla.org/pub/firefox/nightly/2024/01/ 2024-01-01-23-15-40-mozilla-central/}), a local MySQL database, and the Selenium-based~\cite{selenium_webdriver} crawler.
    \item Automated Web Crawl: On each VM, the crawler is launched alongside the MySQL server. The crawler opens Firefox Nightly, installs the Privacy Pioneer extension~\cite{zimmeckEtAlPrivacyPioneer2024}, and visits each URL in the crawl list. During each visit:
    \begin{enumerate}[topsep=3pt,parsep=0pt,itemsep=0pt,leftmargin=*,labelsep=5mm,align=parleft]
        \item Privacy Pioneer monitors HTTP requests in real time.
        \item Requests up to 100,000 characters in length are parsed and classified to detect site connection events.
        \item All identified events are logged into the SQL database through a REST API.
        \item In a separate step, the crawler takes a screenshot of the site to detect cookie banners.
        \item After a fixed observation window of 60 seconds, the crawler proceeds to the next site.
        \item Once crawling is complete, we manually export the collected data for our data analysis. 
    \end{enumerate}
\end{enumerate}

\begin{figure}[H] 
    \centering 
    \includegraphics[width=3.6in]{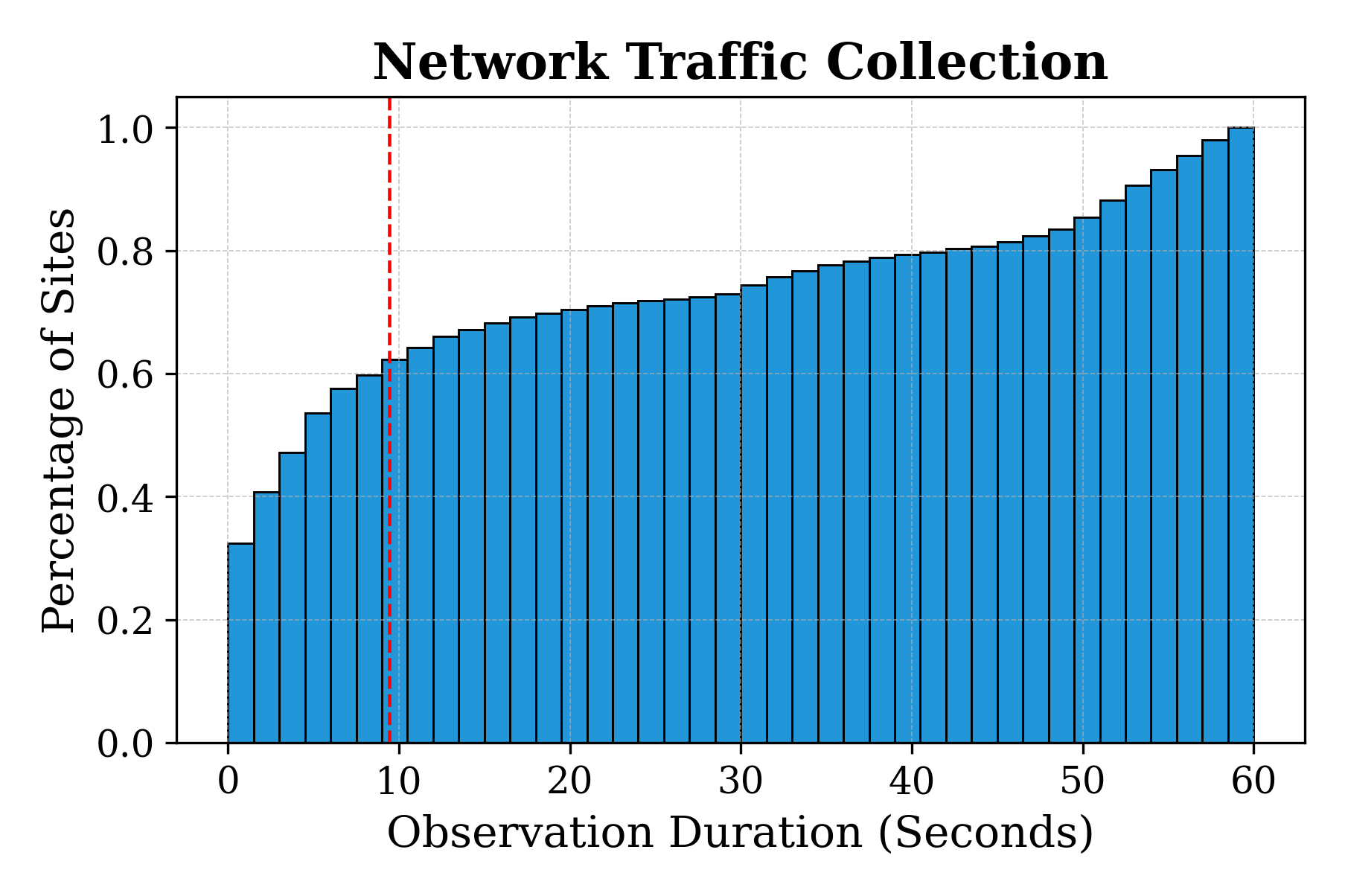} 
    \caption{60\% of sites made no further request after ten seconds of observation. After that time, for the sites in total, the number of requests became much more infrequent before picking up slightly towards the end. Thus, we believe that we captured a meaningful amount of network traffic for most sites.}
    \label{fig:Crawl_Timespan}
\end{figure}

We chose Firefox because our pipeline relies on Privacy Pioneer (\ref{privacy-pioneer}), an existing browser extension only implemented for Firefox. 
Using Firefox therefore ensured compatibility with that prior system and allowed us to preserve methodological continuity with the underlying measurement framework. 
We disabled Firefox's Enhanced Tracking Protection~\cite{mozilla_etp_2024} to avoid browser-level tracker blocking.
To store the collected data we created on each VM a local SQL database~\cite{mysql_installer_download}. 
After the crawler gathers entries from crawled sites the REST API inserts the data into the database.
We performed our crawl from  5--24 January 2025. 
We observed the traffic for each site for 60 seconds (Figure~\ref{fig:Crawl_Timespan}).
Excluding the US (California), crawling the sites on the Common Top 525 and Country-specific Top 525 sites lists took between 19.6 and 21.6 hours for each country (For the US (California) crawling the Common Top 525 sites took 10.9 hours).

\vspace{-16pt}

If a site's SSL/TLS certificate was invalid, improperly configured, or it did not have certificate at all, our crawler would flag the site for an \texttt{InsecureCertificateError} 
 (Table~\ref{tab:insecurecert-error-counts}). 

\vspace{-12pt}
\begin{table}[H]
  \centering
  \small
      \caption{\texttt{InsecureCertificateError} counts and percentages for all sites (All), Country-specific Top 525 sites (Specific), and Common Top 525 sites (Common). For the US (California) the Common Top 525 sites double as Country-specific Top 525 sites.}
      \newcolumntype{C}{>{\centering\arraybackslash}X} 
    \begin{tabularx}{\textwidth}{CCC}
    \toprule
    \multirow{2}{*}{\textbf{Country}} & \textbf{Error Count} & \textbf{Error \%} \\
                     & \textbf{All (Specific, Common)} & \textbf{All (Specific, Common)} \\
    \midrule
    South Africa & 12 (10, 2) & 1.1\% (1.9\%, 0.4\%)\\
    Singapore & 8 (6, 2) & 0.8\% (1.1\%, 0.4\%) \\
    India & 6 (5, 1) & 0.6\% (1.0\%, 0.2\%) \\
    Spain & 6 (4, 2) & 0.6\% (0.8\%, 0.4\%) \\
    Brazil & 5 (4, 1) & 0.5\% (0.8\%, 0.2\%) \\
    Canada & 4 (4, 0) & 0.4\% (0.8\%, 0.0\%) \\
    South Korea & 4 (3, 1) & 0.4\% (0.6\%, 0.2\%) \\
    US (California) & 2 (2, 2) & 0.4\% (0.4\%, 0.4\%) \\
    Australia & 2 (1, 1) & 0.2\% (0.2\%, 0.2\%) \\
    Germany & 1 (1, 0) & 0.1\% (0.2\%, 0.0\%) \\
    \bottomrule
    \end{tabularx}
    \label{tab:insecurecert-error-counts}
\end{table}


\subsection{Tracker Categorization with Privacy Pioneer} 
\label{privacy-pioneer}

We perform the analysis of whether a site connects to a third-party tracker with Privacy Pioneer~\cite{zimmeckEtAlPrivacyPioneer2024}, an open-source browser extension that we install in Firefox Nightly.
The categorization into advertising, analytics, and social categories cover, respectively, third-party ad tracking, site visitor measurement (e.g., of users' geographic regions), and social media integrations. 
Privacy Pioneer identifies whether a site belongs to a particular category based on rule-based matching of request URLs against the Disconnect Tracker Protection lists~\cite{disconnect_tracker_protection}, from which it also retrieves the ``parent company'' of the domain. 
We use the parent-company labels to aggregate individual tracker domains into company-level entities in our analysis, for example, for comparing the most prevalent parent companies across countries and when reporting company-level tracker rates (Section~\ref{sec:Impact of Cookie Banner Consent and Inactivity}) (Privacy Pioneer has additional analysis functionality, for example, identifying if a site collects location data or performs browser fingerprinting, which, however, we do not consider for purposes of our study here). 



\subsection{Cookie Banner Identification and Interaction}

To identify sites with cookie banners our crawler automatically saves screenshots for all sites it visits. 
We then filter for the presence of cookie banners by manually checking all screenshots.
We consider a site to have a cookie banner if it presented a visible notice that informed users about data collection or sharing practices or explicitly requested user consent for such. 
As long as the site provided some on-screen notice of data collection, sharing, or cookie settings we counted it as a cookie banner.
Our crawler only took screenshots of each site without interacting with any cookie banner.
In EU jurisdictions, Germany and Spain, this scenario represents a refusal to consent while, for example, in the US (California) it represents being opted in.
For a random sample of 36 sites from the Common Top 525 sites in three countries, we also manually interacted with the cookie banners to compare the effects of consent and inactivity (Section~\ref{sec:Impact of Cookie Banner Consent and Inactivity}).

\subsection{Limitations}

Our findings should be interpreted in light of various limitations.
First, our VM approach provides a controlled environment but cannot fully represent the local tracking experiences of real users.
Further, except for our manual interaction with the cookie banners for a sample of sites in our dataset, our methodology only captures data from initial site loads without user interaction, which may trigger additional trackers or may make sites behave differently than what we observed. 
Our evaluation is based on crawling a limited set of sites for a limited set of countries using top-level domains as a heuristic for identifying country-specific sites.
We cannot determine whether a country's privacy law is causing the tracking behavior that we observe but only whether the two correlate.
Our approach measures tracker connections rather than data collection. 
Some of these connections may not be actually privacy-invasive, for example, because the incoming data and its related logs are immediately deleted by the receiving site.
However, we can be sure that data was not sent if there was no connection.
Our evaluation is based on Privacy Pioneer's rule-based URL classification that matches request URLs against the Disconnect Tracker Protection block lists~\cite{disconnect_tracker_protection}.
We rely on the correctness of the list's tracker categorization and do not capture tracker connections beyond the list.
We also consider server-side connections beyond the scope of our study.

\section{Additional Figures}
\label{Appendix:Figures_n_Tables}

\vspace{-13pt}
\begin{figure}[H]
    \centering
    \includegraphics[width=3.1in]{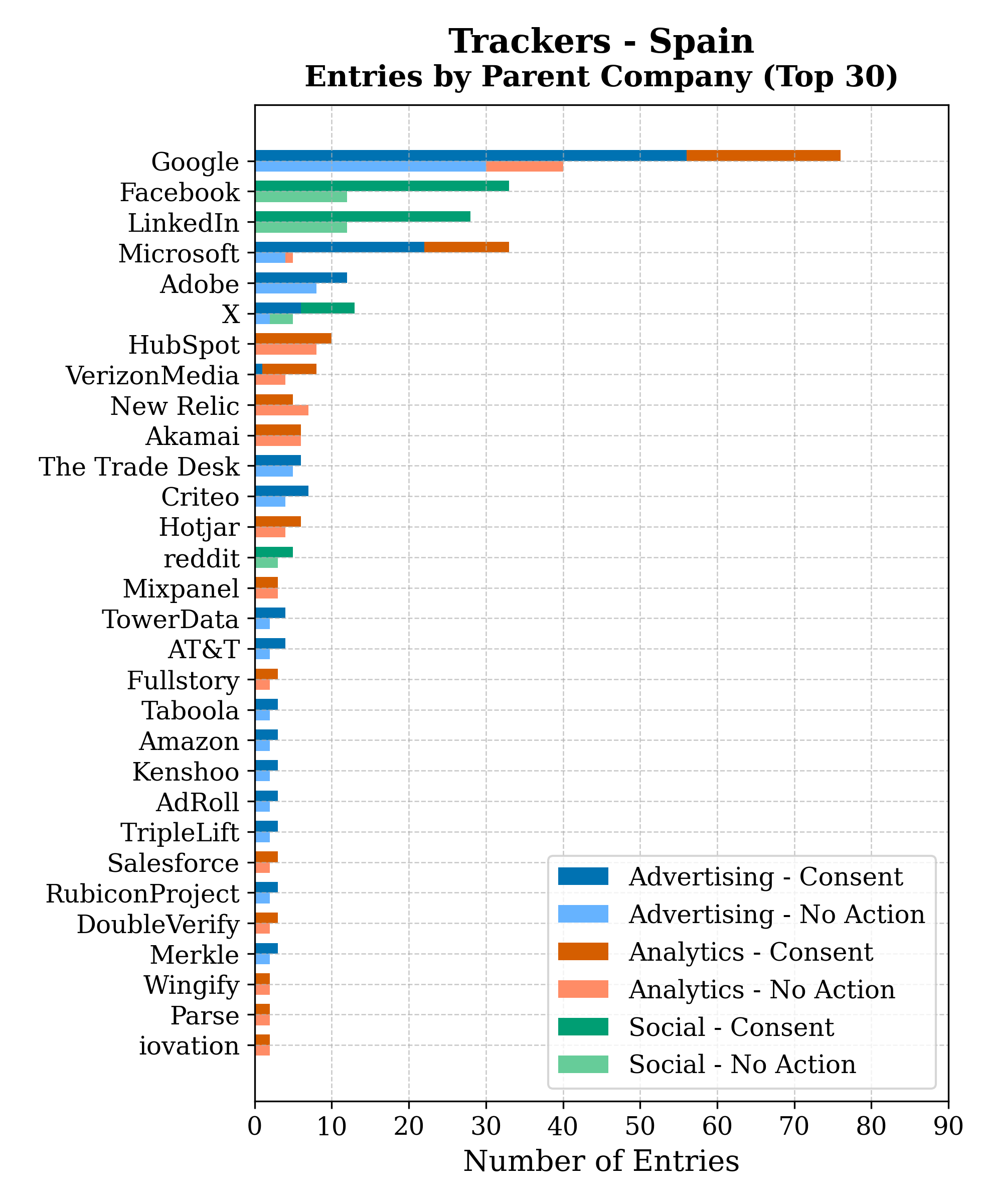}
    \caption{For Spain the number of network connections upon not interacting with cookie banners is often reduced by at least half compared to consenting to the data collection.}
    \label{fig:Sampled_50_Sites_1stPartyExcluded_errorsExcluded_Spain}
\end{figure}
\vspace{-26pt}

\begin{figure}[H]
  \centering
  \includegraphics[width=0.85\textwidth]{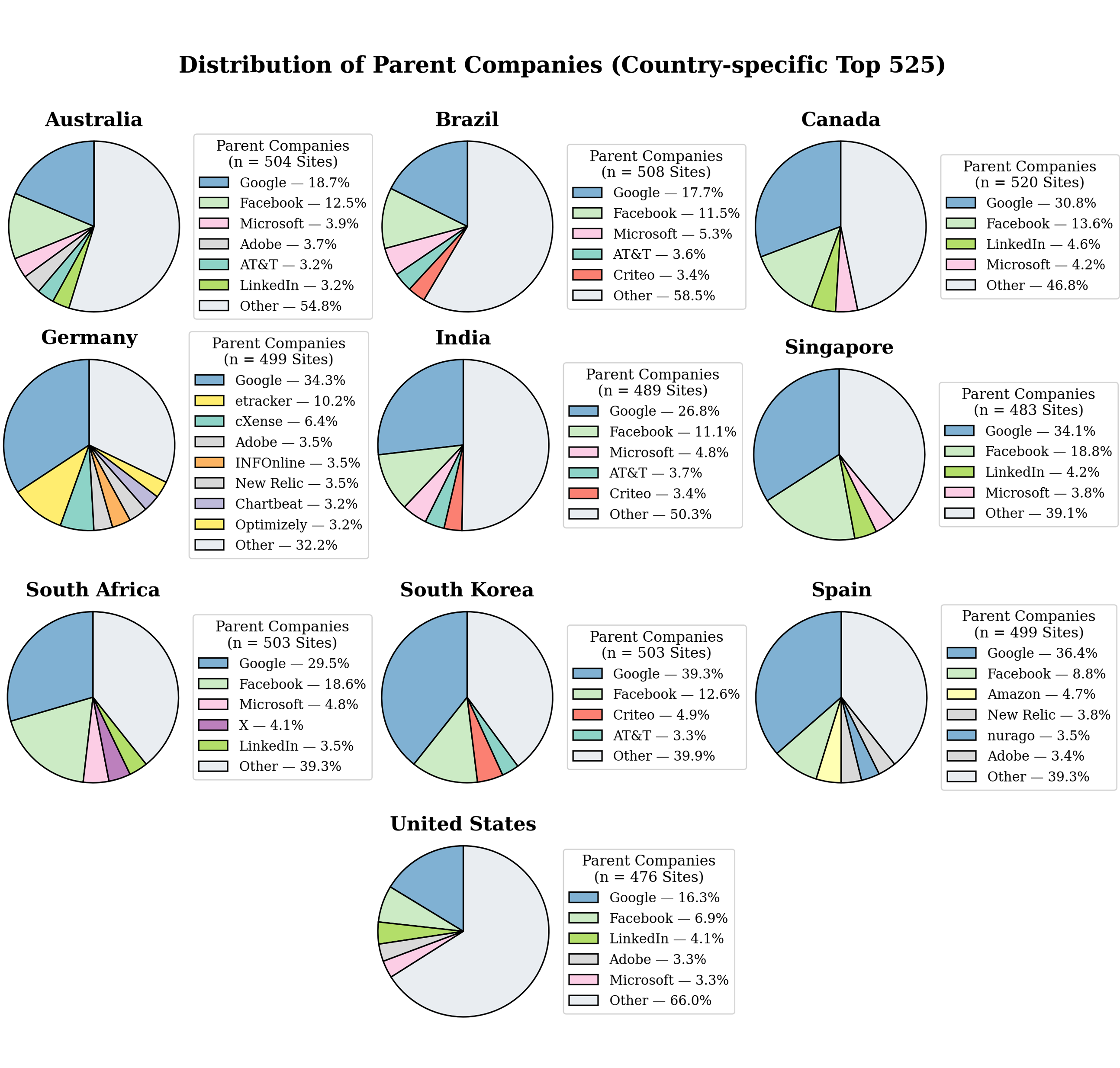}
  \caption{Parent companies for the country-specific Top 525 sites.}
  \label{fig:parent-company-country-top-525}
\end{figure}

\vspace{-16pt}

	\small
	\bibliographystyle{scilight}
	
	

\end{document}